\documentclass[aps,prd,amsmath,amssymb,lengthcheck,showpacs,floatfix,groupedaddress,superscriptaddress,longbibliography]{revtex4-1}
\usepackage{amsmath}
\usepackage{amsfonts}
\usepackage{amssymb}
\usepackage{graphicx}
\usepackage[english]{babel}
\usepackage{times}
\usepackage{dcolumn}
\usepackage[usenames,dvipsnames]{color}
\usepackage{units}
\usepackage[utf8]{inputenc}
\usepackage{hyperref}
\hypersetup{
  colorlinks   = true,    
  urlcolor     = blue,    
  linkcolor    = blue,    
  citecolor    = green      
}

\usepackage{pdfpages}

\pdfpageattr {/Group << /S /Transparency /I true /CS /DeviceRGB>>} 

\makeatletter
\AtBeginDocument{\let\LS@rot\@undefined} 
\makeatother

\begin{document}

\title{Simple predictors of $T_c$ in superconducting cuprates and the role of interactions between effective Wannier orbitals in the $d-p$ 3-band model}

\author{Jakša Vučičević}

\affiliation{Scientific Computing Laboratory, Center for the Study of
Complex Systems,\\
Institute of Physics Belgrade, University of Belgrade, Pregrevica 118,
11080 Belgrade, Serbia}

\author{Michel Ferrero}

\affiliation{CPHT, CNRS, Ecole Polytechnique, Institut Polytechnique de Paris, Route de Saclay, 91128 Palaiseau, France}
\affiliation{Coll\`ege de France, 11 place Marcelin Berthelot, 75005 Paris, France}

\date{\today}

\begin{abstract}
At optimal doping, different cuprate compounds can exhibit vastly different critical temperatures for superconductivity ($T_c$), ranging from about 20 K to about 135 K. The precise properties of the lattice that determine the magnitude of the $T_c$ are currently unknown.
In this paper, we investigate the dependence of the optimal doping $T_c$ on the parameters of the Emery ($d-p$) model for the CuO$_2$ planes in the cuprates.
We show that the best scaling is obtained not with the parameters of the model written in the real ($d/p$-orbital) space, but rather written in the space of effective Wannier orbitals. In this basis, one obtains a model of three sublattices coupled through all possible 4-point interactions.
We identify multiple predictor variables that fit the experimental $T_c$ to about $\pm4-5$ K and that remarkably depend on the leading attractive coupling constants in the transformed Hamiltonian.
\end{abstract}


\maketitle

\newcommand{\expv}[1]{\langle #1 \rangle}
\newcommand{\tk}{\tilde{\mathbf{k}}}
\newcommand{\ImG}{\mathrm{Im}G}

Finding ways to increase the superconducting critical temperature in cuprate compounds is one of the central goals in condensed matter physics\cite{Malozemoff2005,LeeRMP2006,Guven2011,Norman2011}.
The record $T_c$ remains at about 135 K for already more than two decades \cite{Schilling1993,Keimer2015}(if we only consider systems at atmoshperic pressure \cite{Troyan2021,Drozdov2015}). One of the reasons for the lack of progress is that there is no clear understanding of what to look for in a crystal structure, if one is to identify a high-$T_c$ candidate. Many works focused on how $T_c$ correlates with the tight-binding parameters \cite{Ohta1991,Feiner1992,Raimondi1996,Maier2000,Pavarini2001,Hassan2008,Kent2008,Kancharla2008,Zhou2010,Weber2012,Chen2013,Romer2015,Chen2015,Vucicevic2017,Peng2017,Jiang2019,Qin2020,Zhang2023}.
The role of phonons~\cite{Devereaux1995,Devereaux2004,Johnston2010,Wang2021,Rosenstein2021} and disorder\cite{Fukuzimi1996,Eisaki2004,Hobou2009} have been considered as well. In more recent machine learning approaches\cite{Kim2018,Stanev2018,Lee2021,Wang2023}, a large number of different quantities was considered systematically.
Even though the Coulomb interaction is widely believed to be responsible for superconductivity in the cuprates, no works to our knowledge have attempted to systematically link \emph{ab initio}-computed coupling constants to the experimentally measured $T_c$ for multiple compounds.

Studies so far have mostly looked at correlations between the $T_c$ and the parameters of two kinds of models - single-band and three-band models.
In the single-band picture, the main idea was that longer range hopping ($t'$) frustrates the antiferromagnetic (AFM) correlations, which are believed to act as the pairing glue in the cuprates\cite{Prelovsek2005,Wang2014,Metlitski2010,Onufrieva2009,Onufrieva2012,Vucicevic2017,OMahony2022}.
However, the experimentally observed trend in $T_c(t'/t)$ \cite{Raimondi1996,Pavarini2001,Weber2012} was not reproduced in single-band calculations\cite{Maier2000,Hassan2008,Kancharla2008,Chen2013,Jiang2019,Qin2020,Jiang2021,Zhang2023}, thus suggesting that the single-band models (both Hubbard and $tt'J$) do not capture all the mechanisms that determine the $T_c$ in the cuprates.
In the 3-band $d-p$ (Emery) model picture, some works considered the charge-transfer gap (CTG, the difference in energy between copper $d$ and oxygen $p$ orbitals\cite{Weber2012,Weber2014,OMahony2022}, or defined by the gap in the local spectral function\cite{Kowalski2021,OMahony2022}) as the relevant energy scale that determines the strength of the effective AFM coupling, and thus the $T_c$. At least some trends of how the experimental $T_c$ depends on the tight-binding parameters computed for the Emery model can be reproduced by many-body calculations (see Ref.\onlinecite{Weber2012} and compare to Ref.\onlinecite{Kent2008}). More recently, experimentally observed trends of how $T_c$ depends on the density of holes on the copper and oxygen sites separately\cite{Rybicki2016} was also reproduced in calculations for the $d-p$ model\cite{Kowalski2021}. These findings seem to indicate that the Emery model is more relevant for the description of the $T_c$-magnitude in the cuprates.
However, the attempt\cite{Weber2012} to quantitatively correlate the tight-binding parameters of the Emery model to the experimentally measured $T_c$ yielded only poor fits, with large standard deviation of about 30 K. This still leaves open the question of the practical relevance of the CTG and the Emery model.

In this paper, we show that the experimentally measured $T_c$ can indeed be described by a simple function of 3 Emery-model tight-binding parameters (computed for each compound using \emph{ab initio} methods), with a small standard deviation of about 7 K. Furthermore, we show that the interplay between interaction and geometry plays an essential role, and that even better fits can be obtained if one considers not only the tight-binding parameters, but also the coupling constants. The effective onsite repulsion on copper sites $U_{dd}$ is unlikely to depend strongly on the specifics of the lattice structure;
However, if one transforms the Hamiltonian in such a way that the e-e coupling and the kinetic energy become entangled, the resulting coupling constants can be strongly material dependent. By using one such (exact) transformation, we formally obtain a model of 3 separate square lattices, coupled through all possible 4-point interactions between two electrons. Among the coupling constants, some are positive (repulsive), and some are negative (attractive).
We find that the experimentally measured $T_c$ can be fit to within about 5 K, by using a linear function of only 2 parameters of our transformed Hamiltonian, one of them being the leading attractive interaction. We explore the correlations of $T_c$ with the parameters of our transformed model in a systematic and unbiased way. Our results indicate the presence of additional pairing (or pair-breaking) mechanisms in the cuprates, which might strongly affect the magnitude of $T_c$. These mechanisms do not have a simple interpretation in terms of the $d$ and $p$ orbitals, but are apparently related to density-assisted hopping processes between certain spatially extended states, as captured by the Hamiltonian terms in our transformed model. This is particularly interesting in the view of the recent publication of \emph{Jiang et al.}\cite{Jiang2023} which showed in a many-body calculation that such coupling terms can indeed strongly affect the $T_c$.

\begin{figure}[t!]
\centering
\includegraphics[width=1.0\columnwidth, trim=0cm 0 0 0, clip]{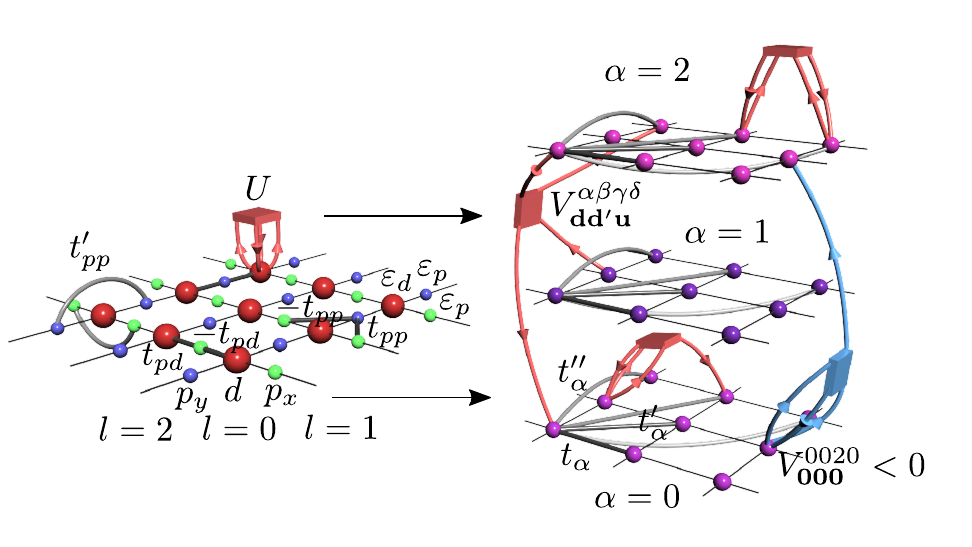}
%
\caption{ Illustration of the Hamiltonian transformation.
A lattice model of $d$ and $p$ orbitals with only local density-density interactions on the $d$-orbitals, is exactly transformed into a model of three square lattices with longer range hoppings, and a zoo of 4-point interactions, some of which are attractive.
}
%
\label{fig:illustration}
\end{figure}

\emph{Model.} The Emery model~\cite{Emery1987,Weber2012,Kowalski2021} (illustrated on Fig.~\ref{fig:illustration}) has a unit cell that contains a copper $d$-orbital and two oxygen $p$-orbitals (denoted with $l=0,1,2\equiv d,p_x,p_y$). The $d$-orbitals form a square lattice, and the $p$-orbitals are found in between the neighboring $d$-orbitals. The hopping amplitude between $d$ and $p$ orbitals is $\pm t_{pd}$, depending on the direction,
and, similarly, between the $p_x$ and $p_y$ orbitals the hopping amplitude is $\pm t_{pp}$. The hopping between the nearest $p_x$ ($p_y$) orbitals is $t'_{pp}$.

The non-interacting part of the Hamiltonian can be diagonalized by switching to the basis of appropriate Bloch waves $d^\dagger_{\alpha,\sigma,\mathbf{k}}|0\rangle$ (see Supplemental Material, SM \cite{SuppMat})
\begin{equation}\label{eq:H0_diagonal}
 \hat{H}_0 = \sum_{\sigma,\alpha,\mathbf{k}} E_{\alpha,\mathbf{k}} d^\dagger_{\alpha,\sigma,\mathbf{k}} d_{\alpha,\sigma,\mathbf{k}},
\end{equation}
where $\alpha=0,1,2$ enumerates the eigenbands in the order of ascending energy, and $\mathbf{k}$ is a wave-vector in the first Brillouin zone. The spin projection ($\uparrow,\downarrow$) is denoted $\sigma$.

In terms of the original local orbitals (denoted $l$), the interacting part of the Hamiltonian, as considered in Refs.\cite{Weber2012} and \cite{Kowalski2021} can be written in two spin-symmetric ways
\begin{eqnarray}\label{eq:quartic_part}
 &&\hat{H}_\mathrm{int} = \frac{1}{2}\sum_{l,\sigma,\mathbf{r}}U_l c^\dagger_{l,\sigma,\mathbf{r}}c_{l,\sigma,\mathbf{r}}c^\dagger_{l,\bar{\sigma},\mathbf{r}}c_{l,\bar{\sigma},\mathbf{r}} \\ \nonumber
 &&= \frac{1}{2}\sum_{l,\sigma\sigma',\mathbf{r}} U_l c^\dagger_{l,\sigma,\mathbf{r}}c_{l,\sigma,\mathbf{r}}c^\dagger_{l,\sigma',\mathbf{r}}c_{l,\sigma',\mathbf{r}}- \sum_{l,\sigma,\mathbf{r}}\frac{U_l}{2} c^\dagger_{l,\sigma,\mathbf{r}}c_{l,\sigma,\mathbf{r}}\\ \label{eq:quartic_part_ss}
\end{eqnarray}
with $U_l = U\delta_{l,0}$, and $\bar\sigma$ denotes the spin projection opposite of $\sigma$. The real-space position of the unit cell is denoted $\mathbf{r}$. The expressions Eq.\ref{eq:quartic_part} and Eq.\ref{eq:quartic_part_ss} are equivalent. However, the choice of one or the other will make a difference for the final form of the Hamiltonian that we reach, following our (exact) transformation: the values of the constants in front of different Hamiltonian terms that we obtain, as well as their physical meaning, will depend on how we initially formulate the interacting part. The quadratic term in Eq.~\ref{eq:quartic_part_ss} will be absorbed in the non-interacting part, and will amount to a shift $\varepsilon_d \rightarrow \varepsilon_d-U/2$. More importantly, only the choice Eq.\ref{eq:quartic_part_ss} will yield a formulation with a spin-rotational symmetry.
We will refer to the formulation based on Eq.\ref{eq:quartic_part} (Eq.\ref{eq:quartic_part_ss}) as the Model A (Model B).

We now rewrite the entire Hamiltonian in the eigenbasis of the non-interacting part, and then further perform the (inverse) Fourier transformation: We express the Hamiltonian in terms of the operators
$
  d^\dagger_{\alpha,\sigma,\mathbf{r}} = \frac{1}{\sqrt{N}}\sum_{\mathbf{k}} e^{-i\mathbf{k}\cdot\mathbf{r}} d^\dagger_{\alpha,\sigma,\mathbf{k}}
$.
There is a phase ambiguity associated with the definition of the operators $d^\dagger_{\alpha,\sigma,\mathbf{k}}$ \cite{Marzari2012}
which we discuss in more detail in SM \cite{SuppMat}. The choice of the phase we make ensures that, in
the final form of the Hamiltoanian, all hopping amplitudes and coupling constants are purely real and are consistent with the symmetries of the original lattice. We obtain
\begin{eqnarray}\label{eq:transformed_H}
  &&\hat{H} = \sum_{\sigma,\alpha,\mathbf{r}\mathbf{d}} t_{\alpha,\mathbf{d}}d^\dagger_{\alpha,\sigma,\mathbf{r}}d_{\alpha,\sigma,\mathbf{r}+\mathbf{d}} \\ \nonumber
  &&\;+ \frac{1}{2}\sum_{\substack{\sigma\sigma',\alpha\beta\gamma\delta \\ \mathbf{r}\mathbf{d}\mathbf{d}'\mathbf{u}}} V^{\alpha\beta\gamma\delta}_{\mathbf{d}\mathbf{d}'\mathbf{u}} d^\dagger_{\alpha,\sigma,\mathbf{r}}d_{\beta,\sigma,\mathbf{r}+\mathbf{d}}d^\dagger_{\gamma,\sigma',\mathbf{r}+\mathbf{u}-\mathbf{d}'}d_{\delta,\sigma',\mathbf{r}+\mathbf{u}}
\end{eqnarray}
where $t_{\alpha,\mathbf{d}}$ is the inverse Fourier transform of $E_{\alpha,\mathbf{k}}$, and it has full square-lattice symmetry. For the precise definition of the coupling constants $V^{\alpha\beta\gamma\delta}_{\mathbf{d}\mathbf{d}'\mathbf{u}}$ see SM \cite{SuppMat}.

The transition from Eqs.~\ref{eq:H0_diagonal} and \ref{eq:quartic_part_ss} to Eq.\ref{eq:transformed_H} is exact, and is illustrated in Fig.~\ref{fig:illustration}.
Starting from Eqs.~\ref{eq:H0_diagonal} and \ref{eq:quartic_part} instead, the only formal difference is the absence of the $\sigma=\sigma'$ terms in the interacting part in Eq.~\ref{eq:transformed_H}, but $t_{\alpha,\mathbf{d}}$ and $V^{\alpha\beta\gamma\delta}_{\mathbf{d}\mathbf{d}'\mathbf{u}}$ values will also be different (due to the absence of the shift $\varepsilon_d \rightarrow \varepsilon_d-U/2$).
%

\emph{Dataset.}  We revisit the dataset compiled by \emph{Weber et al.}~\cite{Weber2012,Weber2012SuppMat}, where Emery model parameters were evaluated from density functional theory (DFT) band-structures for 16 different cuprates (stoichiometric, parent compounds). For two 3-layer compounds, parameters were computed separately for the inner and outer layers, which makes the total number of data points in the dataset 18. The data includes the 4 parameters of the quadratic part of the Hamiltonian written in real-space (onsite energies and hopping amplitudes), as well as the ratio of the next-nearest and the nearest neighbor hoppings $t'/t$ in an effective single-band model that \emph{Weber et al.} derived based on the $d-p$ model parameters. The density-density interaction was only assumed to exist on the $d$-orbitals, and was considered to be the same for all compounds, 8 eV.

\begin{figure}
\centering
\includegraphics[width=0.93\columnwidth, trim=0cm 0 0 0, clip]{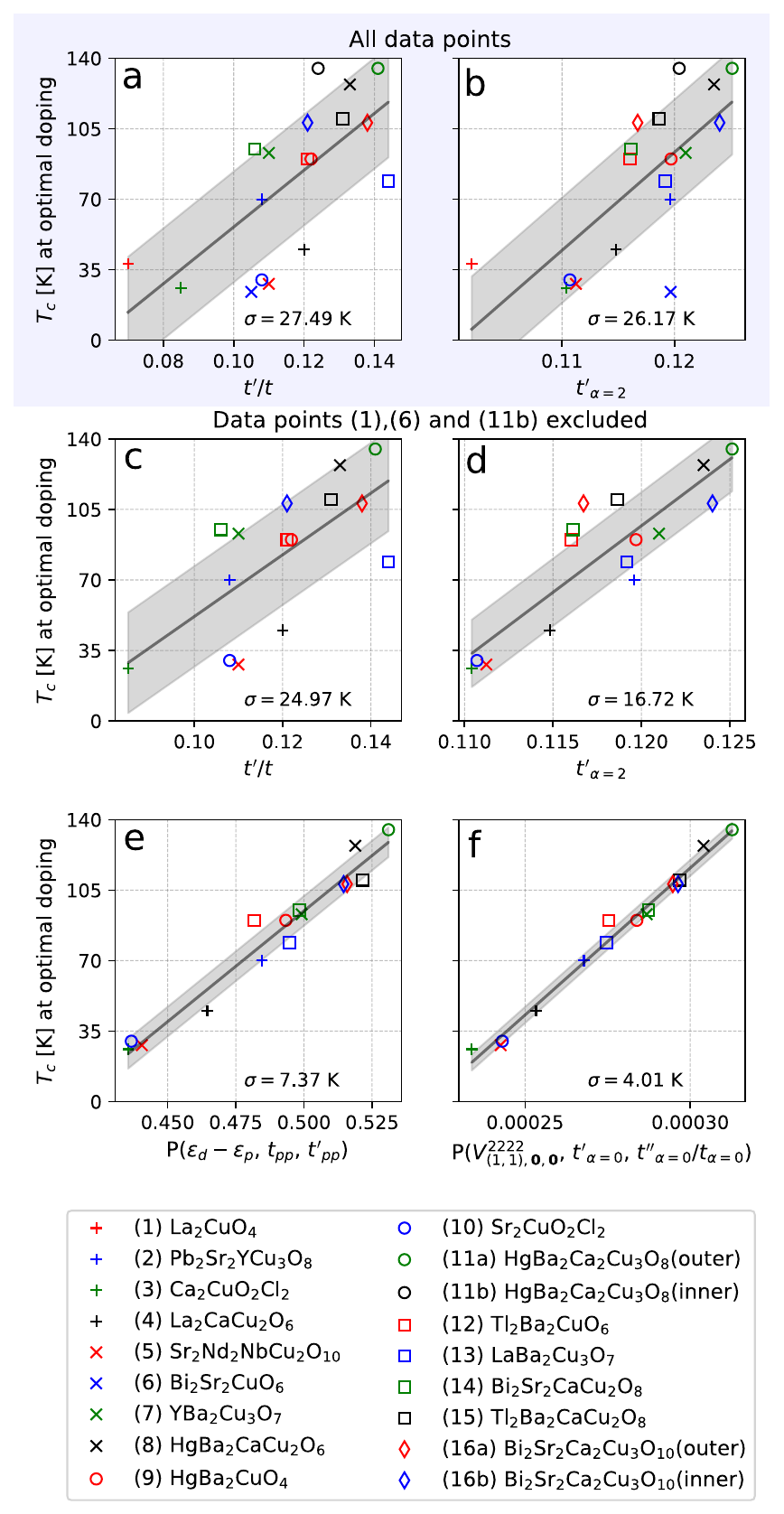}
\caption{ Test of different predictors of $T_c$.
The standard deviation of each fit is denoted $\sigma$.
Black line is the linear fit, the width of the gray shading corresponds to $\pm \sigma$.
}
%
\label{fig:main}
\end{figure}

\emph{Weber et al.} only fitted $T_c$ to individual model parameters. The fits were rather poor (see SM \cite{SuppMat} and Fig.~\ref{fig:main}a). The $T_c$ was found to correlate with $\varepsilon_d-\varepsilon_p$ (the CTG) in the expected way, but only weakly. In our opinion, one should not expect that a single Hamiltonian term controls the $T_c$ in its entirety. One should rather expect a competition (or cooperation) between different processes encoded in the Hamiltonian. Most generally, if the Emery model is correct for the cuprates, the $T_c$ should in general be a single-valued function of \emph{all} the parameters, $T_c(\varepsilon_d-\varepsilon_p, t_{pd},t_{pp},t'_{pp})$. This was not checked in \emph{Weber et al.}, and based on their analyses, one cannot give a clear assessment of the relevance of the Emery model for the cuprates.
We provide such a check on Fig.~\ref{fig:main}e (see also SM \cite{SuppMat}). We demonstrate that a linear combination of three of the Emery model parameters, namely $\epsilon_d-\epsilon_p$, $t_{pp}$ and $t'_{pp}$ is a solid predictor of $T_c$, to within $\pm7.4$K in the whole range of $T_c$, except for three apparent outliers (see the explanations in the next section). The remaining variance of our fit could be attributed to $t_{pd}$, but we find that adding this parameter to the linear combination does not bring much improvement - $T_c$ is not a linear function of $t_{pd}$. The remaining variance could also be due to parameters not included in the Emery model. However, $T_c$ does fit linearly and with an even smaller standard deviation to the parameters of our transformed Hamiltonian, as we show in the following; This presents strong evidence that the Emery model indeed captures the mechanisms that dominantly determine the $T_c$.


\emph{Strategy and results.} For each entry in the \emph{Weber et al.} dataset (given in SM \cite{SuppMat}), we evaluate the dispersions and \emph{all} the parameters of the Hamiltonian in Eq.\ref{eq:transformed_H}.
We then compute from these values about 50 variables that we expect might correlate with $T_c$ 
(these include the bandwidths of each band $D_\alpha$, short-distance hoppings $t_\alpha \equiv t_{\alpha,\mathbf{d}=(1,0)}$, $t'_\alpha \equiv t_{\alpha,\mathbf{d}=(1,1)}$, $t''_\alpha \equiv t_{\alpha,\mathbf{d}=(2,0)}$, as well as various short-distance components and extremal values of $V^{\alpha\beta\gamma\delta}_{\mathbf{d}\mathbf{d}'\mathbf{u}}$).

\begin{table*}[t!]
  \begin{center}
    \caption{Summary of the best predictors of $T_c$. In the first column we restrict to only the original 4 parameters of the model and the $t'/t$ for the effective single-band model computed in \emph{Weber et al.} In the second and third columns we include the paramters of the models A and B, respectively, and variables computed from those parameters. }
    \label{tab:summary}
    \begin{tabular}
      {c||c|c||c|c||c|c}
      category & Original parameters and $t'/t$ & $\sigma$ [K] & Model A & $\sigma$ [K] & Model B & $\sigma$ [K]\\
      \hline
      1-param. best & $t'/t$ & 24.97 & $t'_{\alpha=2}$ & 21.70 & $t'_{\alpha=2}$ & 16.72 \\
      2-param. best & $\varepsilon_d-\varepsilon_p$, $t_{pp}$ & 12.67 & $t''_{\alpha=1}$, $\mathrm{min}V^{\alpha\beta\gamma\delta}_{\mathbf{d}\mathbf{d}'\mathbf{u}}$ & 5.42 & $t'_{\alpha=1}$, $V^{0000}_{(1,1),(-1,-1),\mathbf{0}}$ & 5.74 \\
      3-param. best & $\varepsilon_d-\varepsilon_p$, $t_{pp}$, $t'_{pp}$  & 7.37 & $D_{\alpha=1}$, $t''_{\alpha=1}$, $t'_{\alpha=2}$ & 4.47 & $t'_{\alpha=0}$, $\frac{t''_{\alpha=0}}{t_{\alpha=0}}$, $V^{2222}_{(1,1),\mathbf{0},\mathbf{0}}$ & 4.01 \\
      3-param. 2nd best & $\varepsilon_d-\varepsilon_p$, $t_{pp}$, $t'/t$ & 9.74 & $D_{\alpha=1}$, $t''_{\alpha=1}$, $V^{2222}_{\mathbf{0},\mathbf{0},\mathbf{0}}$ & 4.55 & $t'_{\alpha=0}$, $\mathrm{min}V^{0000}_{\mathbf{d}\mathbf{d}'\mathbf{u}}$, $V^{2222}_{(1,1),\mathbf{0},\mathbf{0}}$ & 4.03
    \end{tabular}
  \end{center}
\end{table*}

We first look at the correlation with the $T_c$ of each individual variable, by doing a linear fit and estimating the standard deviation, $\sigma$.
We find that the best predictor is $t'_{\alpha=2}$ (in Model B formulation), yielding a fit with $\sigma = 26.2$ K.
This is slightly better than the $t'/t$ for the effective single band put forward by \emph{Weber et al.}, but $t'/t$ is, indeed, a close second with $\sigma=27.5$ K (see Fig.~\ref{fig:main} top row). We readily see that the data points (1), (6) and (11b) are outliers for both of the best predictors. In our other attempts at fitting the $T_c$, these 3 points were consistently presenting a limiting factor in obtaining a small $\sigma$. Both points (1) and (6) have a very low $T_c$ - the point (6) has even the lowest $T_c$ (it was also found to be an outlier in Ref.~\onlinecite{Ohta1991}), while the point $(1)$ is extreme in terms of many of the model parameters, so we exclude both points from further analysis. The point (11b) represents the parameters for the inner layer of a 3-layer material, and it may be that the outer layer paramaters, given by the data point (11a), are more relevant, so we exclude the point (11b), as well. In total, we are left with 15 data points, for 14 different compounds. We then redo the fits with respect to individual parameters, and we see that $\sigma$ for the $t'/t$-fit has dropped to about 25 K, while the $\sigma$ for the $t'_{\alpha=2}$-fit has dropped to $16.7$ K. In our subset of data which excludes the apparent outliers, $t'_{\alpha=2}$ is by far the best single-parameter predictor of $T_c$. This holds even in the case of Model A.

We now construct all possible linear combinations of any two and three variables, $P\big(p_1,p_2[,p_3]\big) = c_1 p_1 + c_2 p_2[+ c_3p_3]$,
and we keep fixed $\sum_i c_i^2 = 1 $.
For each of the $\sim1200$ pairs $(p_1,p_2)$ and $\sim40000$ triplets $(p_1,p_2,p_3)$,
we pinpoint the minimum in $\sigma(\{p_i\}; \{c_i\})$ using the Nelder-Mead algorithm.
We then rank different pairs and triplets according to the minimum std. dev. that we can obtain, $\min_{\{c_i\}} \sigma(\{p_i\}; \{c_i\})$.
Finaly, we count the number of times each variable appears in the top 100 triplets, to gain insight into which parameters might be most relevant. Our results are summarized in Table~\ref{tab:summary} (see also SM \cite{SuppMat}).

We observe a general trend in our results, regardless of the choice of the formulation of the interaction part (Eq.\ref{eq:quartic_part} or Eq.\ref{eq:quartic_part_ss}) - good predictors are the linear combinations of a hopping amplitude and one or two coupling constants, in many cases the attractive ones, and in most cases those acting within or between the bands $\alpha=0$ and $\alpha=2$, which are precisely the bands having an appreciable amount of $d$-character.

The best two-parameter predictor we find is the linear combination of the overall most attractive component of $V^{\alpha\beta\gamma\delta}_{\mathbf{d}\mathbf{d}'\mathbf{u}}$ and the hopping amplitude $t''_{\alpha=1}$ (obtained in the Model A formulation), yielding $\sigma=5.4$ K. The most attractive component in both Model A and B formulations is the local density-assisted hybridization from band $\alpha=0$ to band $\alpha=2$, $V^{0020}_{\mathbf{0},\mathbf{0},\mathbf{0}}$.

The best result that we have obtained in our unbiased search is given in Fig.~\ref{fig:main}f. A linear combination of $V^{2222}_{(1,1),\mathbf{0},\mathbf{0}}$, $t'_{\alpha=0}$ and $t''_{\alpha=0}/t_{\alpha=0}$, obtained in Model B, yields a fit of $T_c$ with $\sigma=4.01$ K.
The coupling constant $V^{2222}_{(1,1),\mathbf{0},\mathbf{0}}$ is negative, and correponds to an assisted hopping term in the $\alpha=2$ band, say $n_{\alpha,\uparrow,\mathbf{r}} d^\dagger_{\alpha,\downarrow,\mathbf{r}} d_{\alpha,\downarrow,\mathbf{r}+(1,1)}$ (similar to the terms considered in \emph{Jiang et al.}).
The parameters $V^{2222}_{(1,1),\mathbf{0},\mathbf{0}}$, $t'_{\alpha=0}$ appear the most times in the top 100 3-parameter predictors based on the Model B, in total 65 times. 
It is interesting that $t''_{\alpha=0}/t_{\alpha=0}$ correlates closely with
$V^{0000}_{(2,0),\mathbf{0},\mathbf{0}}$, which is, at the same time, the most attractive interaction in the $\alpha=0$ band. Indeed, the linear combination of $V^{2222}_{(1,1),\mathbf{0},\mathbf{0}}$, $t'_{\alpha=0}$ and $\mathrm{min}V^{0000}_{\mathbf{d}\mathbf{d}'\mathbf{u}}$ is our close second best result, with $\sigma=4.03$.

Finally, we find that the local density-density interaction in the $\alpha=2$ band, $V^{2222}_{\mathbf{0}\mathbf{0}\mathbf{0}}$, might be very relevant.
In the Model A, it appears the most times in the top 100 3-parameter predictors, and in the Model B it is in this sense ranked 6th.
In all linear combinations in which it appears, $V^{2222}_{\mathbf{0}\mathbf{0}\mathbf{0}}$ enters with a negative coefficient. Intuitively, a weaker local repulsion could mean a higher $T_c$. The best single-parameter predictor, $t'_{\alpha=2}$, indeed, highly (anti-)correlates with $V^{2222}_{\mathbf{0}\mathbf{0}\mathbf{0}}$ (see Table~\ref{tab:summary} and SM \cite{SuppMat} for details).

\emph{Discussion and prospects for future work.}
Our results provide strong evidence that the Emery model well captures the mechanisms that determine the magnitude of $T_c$ in the cuprates.
We identify multiple terms in the Hamiltonian which appear particularly relevant for the $T_c$, and propose that these correspond to additional pairing and pair-breaking mechanisms that are in competition. These processes can be understood only in terms of the spatially extended, effective Wannier orbitals in the Emery model, which were not considered in earlier works.

In addition, we obtain a large set of predictor variables that can be computed cheaply, and thus used practically in high-throughput~\cite{Lebegue2013,Herper2017,Kumar2023} searches for novel high-$T_c$ candidate structures.
For practical use, the main question is whether the simple relation between $T_c$ and our predictor variables holds outside the region of the parameter-space that is covered by the \emph{Weber et al.} data points. The best strategy is then to look at crystal structures inside or close to that region, and focus on points for which multiple predictors agree. 
We have scanned the parameter space, and we find a case where each of the 4 parameters of the Emery model is inside the range of values for the existing cuprates, and for which our top 100 predictor variables (based on Model B) predict $T_c \approx 195\pm 5$ K. Going only slightly away from the range of Emery model parameters covered by the data points,
we find cases which correspond to $T_c$ of even more than 250 K (see SM \cite{SuppMat}for details).

As was the case with previous similar works, the main limitation of our approach lies in the ambiguity of the DFT calculations~\cite{Lejaeghere2016,Aryasetiawan2022} and the downfolding procedures\cite{Hansmann2014,Jiang2023},
especially when it comes to the choice and computation of Coulomb tensor elements; our work ultimately highlights the necessity of a careful and systematic work in that direction.

\begin{acknowledgments}
We acknowledge useful discussions with A.-M.~S. Tremblay and Antoine Georges. We acknowledge contributions from Bogdan Rajkov and Sidhartha Dash in the early stages of this work. Computations were performed on the PARADOX supercomputing facility (Scientific Computing Laboratory, Center for the Study of Complex
Systems, Institute of Physics Belgrade). J.~V. acknowledges funding provided by the Institute of Physics Belgrade, through the grant by the Ministry of Science, Technological Development and Innovation of the Republic of Serbia. J.~V. acknowledges funding by the European Research Council, grant ERC-2022-StG: 101076100.
\end{acknowledgments}

\bibliography{refs}
\newpage
\newpage
\newpage
\newpage
\begin{widetext}
\newpage


\section*{Supplementary material}

Jakša Vučičević[1] and Michel Ferrero[2,3]

[1]*Scientific Computing Laboratory, Center for the Study of Complex Systems,
Institute of Physics Belgrade, University of Belgrade, Pregrevica 118, 11080 Belgrade, Serbia*
[2]*CPHT, CNRS, Ecole Polytechnique, Institut Polytechnique de Paris, Route de Saclay, 91128 Palaiseau, France*
[3]*Collège de France, 11 place Marcelin Berthelot, 75005 Paris, France*
(Dated: February 2, 2024)

## I. MODEL HAMILTONIAN, TRANSFORMATION TO WANNIER SPACE, AND SYMMETRIES

### A. Non-interacting part

We start from the Hamiltonian written in the basis of $d, p_x, p_y$-orbitals (denoted by $l = 0, 1, 2$, respectively). For the sake of convenience, we define a row-vector of creation operators

$$\mathbf{\Psi}^\dagger_{\sigma,\mathbf{r}} = (c^\dagger_{l=0,\sigma,\mathbf{r}}, c^\dagger_{l=1,\sigma,\mathbf{r}}, c^\dagger_{l=2,\sigma,\mathbf{r}}) \tag{1}$$

where $\mathbf{r}$ denotes the position of the unit cell (set to the position of the $d$-orbital within the unit cell) and $\sigma$ is the spin projection. The non-interacting part of the Hamiltonian reads

$$\hat{H} = \sum_{\sigma,\mathbf{rr}'} \mathbf{\Psi}^\dagger_{\sigma,\mathbf{r}} \mathbf{h}_{\mathbf{rr}'} \mathbf{\Psi}_{\sigma,\mathbf{r}'} \tag{2}$$

where

$$\mathbf{h}_{\mathbf{rr}'} = \begin{pmatrix} \varepsilon_d \delta_{\mathbf{rr}'} & t_{pd}(\delta_{\mathbf{rr}'} - \delta_{\mathbf{r}+\mathbf{e}_x,\mathbf{r}'}) & t_{pd}(\delta_{\mathbf{rr}'} - \delta_{\mathbf{r}+\mathbf{e}_y,\mathbf{r}'}) \\ t_{pd}(\delta_{\mathbf{rr}'} - \delta_{\mathbf{r}-\mathbf{e}_x,\mathbf{r}'}) & \varepsilon_p \delta_{\mathbf{rr}'} + t'_{pp}(\delta_{\mathbf{r}+\mathbf{e}_x,\mathbf{r}'} + \delta_{\mathbf{r}-\mathbf{e}_x,\mathbf{r}'}) & t_{pp}(\delta_{\mathbf{rr}'} - \delta_{\mathbf{r}+\mathbf{e}_y,\mathbf{r}'} - \delta_{\mathbf{r}-\mathbf{e}_x,\mathbf{r}'} + \delta_{\mathbf{r}-\mathbf{e}_x+\mathbf{e}_y,\mathbf{r}'}) \\ t_{pd}(\delta_{\mathbf{rr}'} - \delta_{\mathbf{r}-\mathbf{e}_y,\mathbf{r}'}) & t_{pp}(\delta_{\mathbf{rr}'} - \delta_{\mathbf{r}-\mathbf{e}_y,\mathbf{r}'} - \delta_{\mathbf{r}+\mathbf{e}_x,\mathbf{r}'} + \delta_{\mathbf{r}+\mathbf{e}_x-\mathbf{e}_y,\mathbf{r}'}) & \varepsilon_p \delta_{\mathbf{rr}'} + t'_{pp}(\delta_{\mathbf{r}+\mathbf{e}_y,\mathbf{r}'} + \delta_{\mathbf{r}-\mathbf{e}_y,\mathbf{r}'}) \end{pmatrix} \tag{3}$$

and we define the unit vectors $\mathbf{e}_x = (1,0)$ and $\mathbf{e}_y = (0,1)$, in units of the lattice spacing (distance between neigboring Cu nuclei). We can now apply the Fourer fransform over the unit cells to bring the Hamiltonian into a block-diagonal form. This can be done in multiple ways, as there is a gauge freedom of choosing a $\mathbf{k}$-dependent phase in front of $c^\dagger_{l,\sigma,\mathbf{k}}$ operators. We take the following convention:

$$c^\dagger_{l,\sigma,\mathbf{k}} = \frac{\varphi_{l,\mathbf{k}}}{\sqrt{N}} \sum_\mathbf{r} e^{i\mathbf{k}\cdot\mathbf{r}} c^\dagger_{l,\sigma,\mathbf{r}}, \qquad c_{l,\sigma,\mathbf{k}} = \frac{\varphi^*_{l,\mathbf{k}}}{\sqrt{N}} \sum_\mathbf{r} e^{-i\mathbf{k}\cdot\mathbf{r}} c_{l,\sigma,\mathbf{r}} \tag{4}$$

$$c^\dagger_{l,\sigma,\mathbf{r}} = \frac{1}{\sqrt{N}} \sum_\mathbf{k} \varphi^*_{l,\mathbf{k}} e^{-i\mathbf{k}\cdot\mathbf{r}} c^\dagger_{l,\sigma,\mathbf{k}}, \qquad c_{l,\sigma,\mathbf{r}} = \frac{1}{\sqrt{N}} \sum_\mathbf{k} \varphi_{l,\mathbf{k}} e^{i\mathbf{k}\cdot\mathbf{r}} c_{l,\sigma,\mathbf{k}} \tag{5}$$

with

$$\varphi_{l,\mathbf{k}} = \begin{cases} 1, & l = 0 \\ i\exp\left(-i\frac{k_x}{2}\right), & l = 1 \\ -i\exp\left(-i\frac{k_y}{2}\right), & l = 2 \end{cases} \tag{6}$$

This choice corresponds to doing the sum over the actual real-space positions of the given $d, p_x$ or $p_y$ orbitals, which are found at $\mathbf{r}, \mathbf{r} - \mathbf{e}_x/2$ and $\mathbf{r} - \mathbf{e}_y/2$, respectively. The Fourier transformation Eq. 5 of each operator *a priori* leads to the following form

$$\hat{H} = \sum_{\sigma,\mathbf{kk}'} \mathbf{\Psi}^\dagger_{\sigma,\mathbf{k}} \mathbf{h}_{\mathbf{kk}'} \mathbf{\Psi}_{\sigma,\mathbf{k}'}. \tag{7}$$

Note that the sums over $\mathbf{r}, \mathbf{r}'$ have been absorbed into the matrix $\mathbf{h_{kk'}}$, so that it only depends on $\mathbf{k}, \mathbf{k}'$, corresponding to the momenta of the creation and annihilation operators to the left and to the right of the matrix. Let's evaluate each of the matrix elements of $\mathbf{h_{kk'}}$.

$$
\begin{aligned}
[\mathbf{h_{k,k'}}]_{l=0,l'=0} &= \frac{\varepsilon_d \; \varphi_{0,\mathbf{k}} \varphi_{0,\mathbf{k'}}^*}{N} \sum_{\mathbf{rr'}} \delta_{\mathbf{rr'}} e^{i\mathbf{k}\cdot\mathbf{r}} e^{-i\mathbf{k'}\cdot\mathbf{r'}} \\
&= \varepsilon_d \frac{1}{N} \sum_{\mathbf{r}} e^{i(\mathbf{k}-\mathbf{k'})\cdot\mathbf{r}} \\
&= \varepsilon_d \delta_{\mathbf{kk'}}
\end{aligned}
\tag{8}
$$

$$
\begin{aligned}
[\mathbf{h_{k,k'}}]_{l=0,l'=1} &= \frac{t_{pd} \; \varphi_{0,\mathbf{k}} \varphi_{1,\mathbf{k'}}^*}{N} \sum_{\mathbf{rr'}} e^{i\mathbf{k}\cdot\mathbf{r}} e^{-i\mathbf{k'}\cdot\mathbf{r'}} (\delta_{\mathbf{rr'}} - \delta_{\mathbf{r}+\mathbf{e}_x,\mathbf{r'}}) \\
&= t_{pd} \; \varphi_{1,\mathbf{k'}}^* (1 - e^{-ik_x'}) \frac{1}{N} \sum_{\mathbf{r}} e^{i(\mathbf{k}-\mathbf{k'})\cdot\mathbf{r}} \\
&= t_{pd} \; (-i)(e^{ik_x/2} - e^{-ik_x/2}) \delta_{\mathbf{kk'}} \\
&= 2t_{pd} \sin \frac{k_x}{2} \delta_{\mathbf{kk'}}
\end{aligned}
\tag{9}
$$

$$
\begin{aligned}
[\mathbf{h_{k,k'}}]_{l=1,l'=1} &= \varepsilon_p \delta_{\mathbf{kk'}} + t_{pp}' \frac{\varphi_{1,\mathbf{k}} \varphi_{1,\mathbf{k'}}^*}{N} \sum_{\mathbf{rr'}} e^{i\mathbf{k}\cdot\mathbf{r}} e^{-i\mathbf{k'}\cdot\mathbf{r'}} (\delta_{\mathbf{r}-\mathbf{e}_x,\mathbf{r'}} + \delta_{\mathbf{r}+\mathbf{e}_x,\mathbf{r'}}) \\
&= \varepsilon_p \delta_{\mathbf{kk'}} + t_{pp}' (e^{ik_x} + e^{-ik_x'}) \frac{\varphi_{1,\mathbf{k}} \varphi_{1,\mathbf{k'}}^*}{N} \sum_{\mathbf{r}} e^{i(\mathbf{k}-\mathbf{k'})\cdot\mathbf{r}} \\
&= (\varepsilon_p + 2t_{pp}' \cos k_x) \delta_{\mathbf{kk'}}
\end{aligned}
$$

$$
\begin{aligned}
[\mathbf{h_{k,k'}}]_{l=1,l'=2} &= t_{pp} \frac{\varphi_{1,\mathbf{k}} \varphi_{2,\mathbf{k'}}^*}{N} \sum_{\mathbf{rr'}} e^{i\mathbf{k}\cdot\mathbf{r}} e^{-i\mathbf{k'}\cdot\mathbf{r'}} (\delta_{\mathbf{rr'}} - \delta_{\mathbf{r}+\mathbf{e}_y,\mathbf{r'}} - \delta_{\mathbf{r}-\mathbf{e}_x,\mathbf{r'}} + \delta_{\mathbf{r}-\mathbf{e}_x+\mathbf{e}_y,\mathbf{r'}}) \\
&= t_{pp} \; \varphi_{1,\mathbf{k}} \; \varphi_{2,\mathbf{k}}^* (1 - e^{-ik_y} - e^{ik_x} + e^{ik_x - ik_y}) \delta_{\mathbf{kk'}} \\
&= t_{pp} \; ie^{-ik_x/2} ie^{ik_y/2} (1 - e^{-ik_y})(1 - e^{ik_x}) \delta_{\mathbf{kk'}} \\
&= -4t_{pp} \; \sin \frac{k_x}{2} \sin \frac{k_y}{2} \delta_{\mathbf{kk'}}
\end{aligned}
\tag{10}
$$

The rest of the terms can be obtained completely analogously. As all of the terms are proportional to $\delta_{\mathbf{kk'}}$, we can drop the second index in $\mathbf{h_{kk'}}$, so we get

$$
\hat{H}_0 = \sum_{\sigma,\mathbf{k}} \mathbf{\Psi}_{\sigma,\mathbf{k}}^\dagger \mathbf{h_k} \mathbf{\Psi}_{\sigma,\mathbf{k}}
\tag{11}
$$

with $\mathbf{\Psi}_\mathbf{k}^\dagger = (c_{l=0,\sigma,\mathbf{k}}^\dagger, c_{l=1,\sigma,\mathbf{k}}^\dagger, c_{l=2,\sigma,\mathbf{k}}^\dagger)$, and we have

$$
\mathbf{h_k} = \begin{pmatrix} \varepsilon_d & 2t_{pd} \sin \frac{k_x}{2} & -2t_{pd} \sin \frac{k_y}{2} \\ \text{H.c.} & \varepsilon_p + 2t_{pp}' \cos k_x & -4t_{pp} \sin \frac{k_x}{2} \sin \frac{k_y}{2} \\ \text{H.c.} & \text{H.c.} & \varepsilon_p + 2t_{pp}' \cos k_y \end{pmatrix}.
\tag{12}
$$

We can now diagonalize each block of the non-interacting Hamiltonian, to arrive at Eq.1 from the main text:

$$
\hat{H}_0 = \sum_{\sigma,\alpha,\mathbf{k}} E_{\alpha,\mathbf{k}} d_{\alpha,\sigma,\mathbf{k}}^\dagger d_{\alpha,\sigma,\mathbf{k}}
\tag{13}
$$

where $\alpha$ denotes the eigenbands, and the connection with previously defined operators is given in terms of the (transposed) basis-change matrix $\mathbf{A_k}$

$$
d_{\alpha,\sigma,\mathbf{k}}^\dagger = \sum_l [\mathbf{A_k}]_{\alpha,l} c_{l,\sigma,\mathbf{k}}^\dagger, \quad c_{l,\sigma,\mathbf{k}}^\dagger = \sum_\alpha [\mathbf{A_k}^{-1}]_{l,\alpha} d_{\alpha,\sigma,\mathbf{k}}^\dagger
\tag{14}
$$

Here note that we use the transpose of the basis change matrix $\mathbf{P_k^T} \equiv \mathbf{A_k}$ only for the sake of convenience. The basis change is in general performed by $\mathbf{\Psi}_{\sigma,\mathbf{k}}^\dagger \mathbf{h_k} \mathbf{\Psi}_{\sigma,\mathbf{k}} \to (\mathbf{\Psi}_{\sigma,\mathbf{k}}^\dagger \mathbf{P_k})(\mathbf{P_k^{-1}} \mathbf{h_k} \mathbf{P_k})(\mathbf{P_k^{-1}} \mathbf{\Psi}_{\sigma,\mathbf{k}})$, but it must hold $\mathbf{P_k^{-1}} = (\mathbf{P^T})_\mathbf{k}^*$ because one must have $d_{\alpha,\sigma,\mathbf{k}} = (d_{\alpha,\sigma,\mathbf{k}}^\dagger)^\dagger$, and we check that it does.

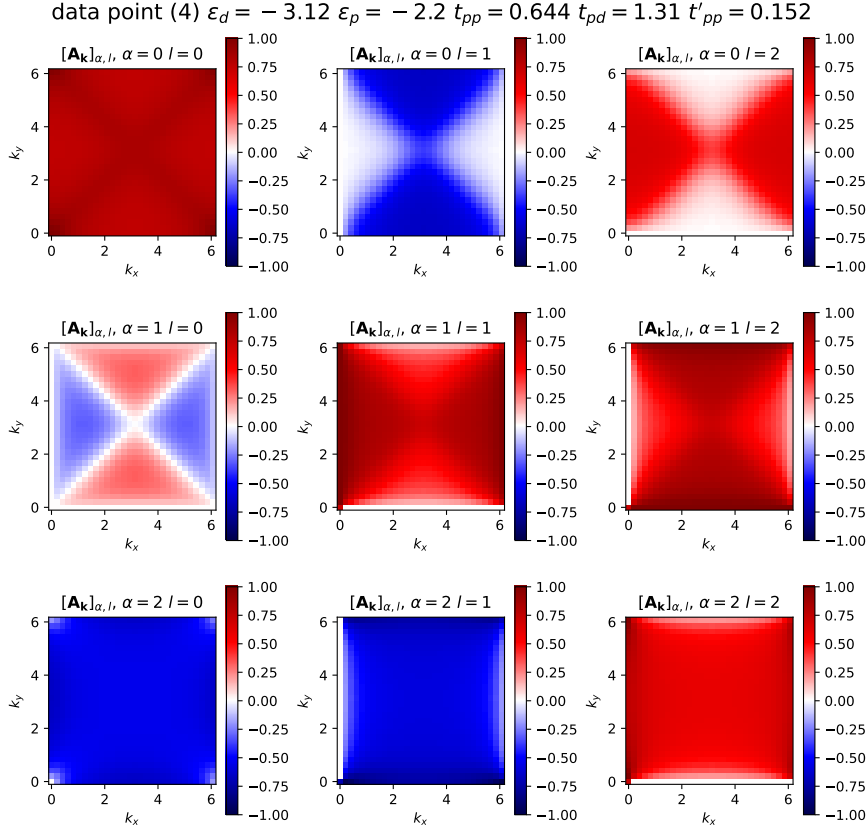

Figure 1. An example of the basis-transformation matrix $\mathbf{A_k}$. Note that on this plot we discretize the Brillouin zone with only 30×30 points, so that the edge cases are more easily inspected. To compute the hopping amplitudes we have used $360 \times 360$ points.

There is a gauge freedom associated with the definition of $d^\dagger_{\alpha,\sigma,\mathbf{k}}$ - eigenstates are defined only up to an overall phase. We make sure that all elements of $[\mathbf{A_k}]_{\alpha,l}$ are purely real, by subtracting any phase from the result we obtain in the following way:

$$[\mathbf{A_k}]_{\alpha,l} \rightarrow [\mathbf{A_k}]_{\alpha,l} e^{-i \arg[\mathbf{A_k}]_{\alpha,l=\alpha}} \tag{15}$$

where the labelling $\alpha = 0, 1, 2$ is ordered by increasing $E_{\alpha,\mathbf{k}}$. In our implementation, special care to ensure smoothness of $[\mathbf{A_k}]_{\alpha,l}$ was only needed for $k_y = 0$ components, where $[\mathbf{A_k}]_{\alpha,l=0} = 0$ and $\arg[\mathbf{A_k}]_{\alpha,l=\alpha}$ is undefined. Additionally, the point $\mathbf{k} = (0,0)$ is special because the eigenstates $\alpha = 1$ and $\alpha = 2$ are degenerate there - then any linear combination of these eigenstates is also an eigenstate. We resolve this by imposing that $[\mathbf{A_k}]_{\alpha,l}$ is smooth passing through $\mathbf{k} = (0,0)$ along the $k_x = k_y$ direction. This gives us

$$[\mathbf{A_{k=(0,0)}}]_{\alpha,l} = \frac{(-1)^{\delta_{\alpha,2}\delta_{l,1}}}{\sqrt{2}} \tag{16}$$

Finally, we can Fourier tranform the operators acting on the eigenbands, and express the entire Hamiltonian in terms of

$$d^\dagger_{\alpha,\sigma,\mathbf{r}} = \frac{1}{\sqrt{N}} \sum_{\mathbf{k}} e^{-i\mathbf{k}\cdot\mathbf{r}} d^\dagger_{\alpha,\sigma,\mathbf{k}}, \quad d_{\alpha,\sigma,\mathbf{r}} = \frac{1}{\sqrt{N}} \sum_{\mathbf{k}} e^{i\mathbf{k}\cdot\mathbf{r}} d_{\alpha,\sigma,\mathbf{k}} \tag{17}$$

with the inverse transformation being

$$d^\dagger_{\alpha,\sigma,\mathbf{k}} = \frac{1}{\sqrt{N}} \sum_{\mathbf{r}} e^{i\mathbf{k}\cdot\mathbf{r}} d^\dagger_{\alpha,\sigma,\mathbf{r}}, \quad d_{\alpha,\sigma,\mathbf{k}} = \frac{1}{\sqrt{N}} \sum_{\mathbf{r}} e^{-i\mathbf{k}\cdot\mathbf{r}} d_{\alpha,\sigma,\mathbf{r}} \tag{18}$$

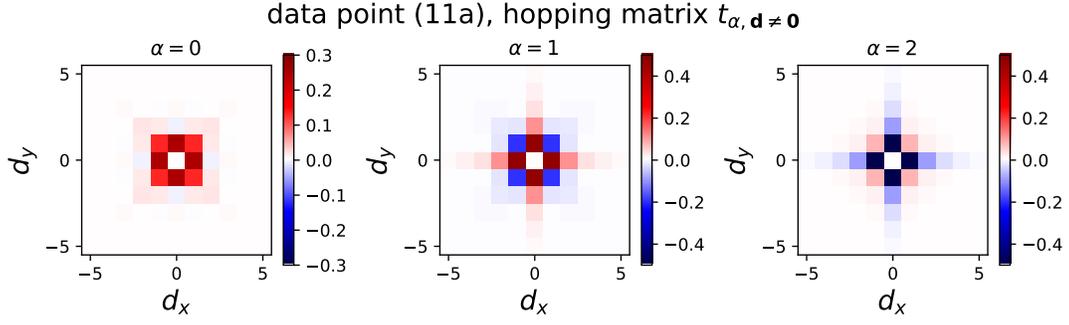

Figure 2. An example of the the hopping amplitudes for the 3 bands, computed with $360 \times 360$ points in the Brillouin zone (Model A formulation).

The non-interacting part now becomes:

$$
\begin{aligned}
\hat{H}_0 &= \sum_{\alpha,\sigma,\mathbf{rr'}} d^\dagger_{\alpha,\sigma,\mathbf{r}} d_{\alpha,\sigma,\mathbf{r'}} \frac{1}{N} \sum_{\mathbf{k}} e^{-i\mathbf{k}\cdot(\mathbf{r'}-\mathbf{r})} E_{\alpha,\mathbf{k}} \\
&= \sum_{\alpha,\sigma,\mathbf{rd}} t_{\alpha,\mathbf{d}} d^\dagger_{\alpha,\sigma,\mathbf{r}} d_{\alpha,\sigma,\mathbf{r+d}}
\end{aligned}
\tag{19}
$$

where

$$
t_{\alpha,\mathbf{d}} = \frac{1}{N} \sum_{\mathbf{k}} e^{-i\mathbf{k}\cdot\mathbf{d}} E_{\alpha,\mathbf{k}}
\tag{20}
$$

is the hopping matrix corresponding to the band $\alpha$. The term $\mathbf{d} = \mathbf{0}$ has the physical meaning of the onsite energy of the Wannier orbital $\alpha$.

The connection between the original $d, p_x, p_y$ orbitals ($|l,\sigma,\mathbf{r}\rangle = c^\dagger_{l,\sigma,\mathbf{r}}|0\rangle$) and the Wannier orbitals ($|\alpha,\sigma,\mathbf{r}\rangle = d^\dagger_{\alpha,\sigma,\mathbf{r}}|0\rangle$) is the following

$$
\begin{aligned}
|\alpha,\sigma,\mathbf{r}\rangle &= \frac{1}{\sqrt{N}} \sum_{\mathbf{k}} e^{-i\mathbf{k}\cdot\mathbf{r}} |\alpha,\sigma,\mathbf{k}\rangle \tag{21} \\
&= \frac{1}{\sqrt{N}} \sum_{\mathbf{k}} e^{-i\mathbf{k}\cdot\mathbf{r}} \sum_l [\mathbf{A_k}]_{\alpha,l} |l,\sigma,\mathbf{k}\rangle \tag{22} \\
&= \frac{1}{\sqrt{N}} \sum_{\mathbf{k}} e^{-i\mathbf{k}\cdot\mathbf{r}} \sum_l [\mathbf{A_k}]_{\alpha,l} \frac{\varphi_{l,\mathbf{k}}}{\sqrt{N}} \sum_{\mathbf{r'}} e^{i\mathbf{k}\cdot\mathbf{r'}} |l,\sigma,\mathbf{r'}\rangle \tag{23} \\
&= \sum_{\mathbf{r'},l} \left( \frac{1}{N} \sum_{\mathbf{k}} e^{-i\mathbf{k}\cdot(\mathbf{r}-\mathbf{r'})} [\mathbf{A_k}]_{\alpha,l} \varphi_{l,\mathbf{k}} \right) |l,\sigma,\mathbf{r'}\rangle \tag{24} \\
&= \sum_{\mathbf{r'},l} F_{\alpha,l,\mathbf{r}-\mathbf{r'}} |l,\sigma,\mathbf{r'}\rangle \tag{25}
\end{aligned}
$$

where $F_{\alpha,l,\mathbf{r}}$ is the inverse Fourier transform of the quantity $[\mathbf{A_k}]_{\alpha,l} \varphi_{l,\mathbf{k}}$.

data point (11a), $|\alpha, \sigma, \mathbf{r} = 0\rangle = \sum_{\mathbf{r}', l} F_{\alpha, l, -\mathbf{r}'} |l, \sigma, \mathbf{r}'\rangle$

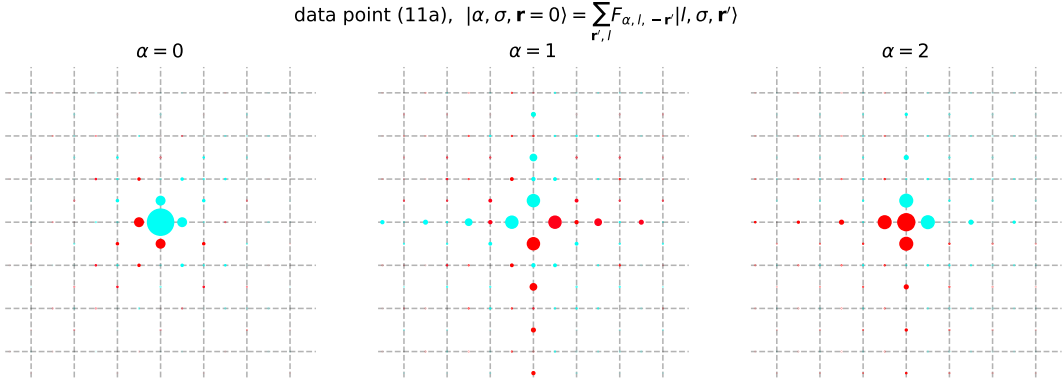

Figure 3. An example of the the Wannier orbitals corresponding to the 3 bands, given in terms of the $d, p_x, p_y$ orbitals (Model A formulation). The area of the point is the amplitude of $F_{\alpha, l, -\mathbf{r}'}$, the color is the phase ($F_{\alpha, l, -\mathbf{r}'}$ is purely real, so the color here denotes the sign).

## B. Interacting part

Let's consider the interacting part (density-density interaction on each orbital $l$):

$$\hat{H}_{\text{int}} = \frac{1}{2} \sum_{l, \sigma, \mathbf{r}} U_l c^{\dagger}_{l, \sigma, \mathbf{r}} c_{l, \sigma, \mathbf{r}} c^{\dagger}_{l, \bar{\sigma}, \mathbf{r}} c_{l, \bar{\sigma}, \mathbf{r}} \tag{26}$$

$$= \frac{1}{2} \sum_{l, \sigma} U_l \sum_{\mathbf{k}_1 \ldots \mathbf{k}_4} \varphi^{*}_{l, \mathbf{k}_1} \varphi_{l, \mathbf{k}_2} \varphi^{*}_{l, \mathbf{k}_3} \varphi_{l, \mathbf{k}_4} c^{\dagger}_{l, \sigma, \mathbf{k}_1} c_{l, \sigma, \mathbf{k}_2} c^{\dagger}_{l, \bar{\sigma}, \mathbf{k}_3} c_{l, \bar{\sigma}, \mathbf{k}_4} \frac{1}{N^2} \sum_{\mathbf{r}} e^{-i(\mathbf{k}_1 - \mathbf{k}_2 + \mathbf{k}_3 - \mathbf{k}_4) \cdot \mathbf{r}} \tag{27}$$

$$= \frac{1}{2N} \sum_{l, \sigma} U_l \sum_{\mathbf{k}\mathbf{k}'\mathbf{q}} \varphi^{*}_{l, \mathbf{k}+\mathbf{q}} \varphi_{l, \mathbf{k}} \varphi^{*}_{l, \mathbf{k}'} \varphi_{l, \mathbf{k}'+\mathbf{q}} c^{\dagger}_{l, \sigma, \mathbf{k}+\mathbf{q}} c_{l, \sigma, \mathbf{k}} c^{\dagger}_{l, \bar{\sigma}, \mathbf{k}'} c_{l, \bar{\sigma}, \mathbf{k}'+\mathbf{q}} \tag{28}$$

We now introduce $\mathcal{F}_{l, \mathbf{k}\mathbf{k}'\mathbf{q}} \equiv \varphi^{*}_{l, \mathbf{k}+\mathbf{q}} \varphi_{l, \mathbf{k}} \varphi^{*}_{l, \mathbf{k}'} \varphi_{l, \mathbf{k}'+\mathbf{q}}$ and transform into the eigenbasis of the non-interacting part:

$$\hat{H}_{\text{int}} = \frac{1}{2N} \sum_{\sigma, \alpha\beta\gamma\delta} \sum_{\mathbf{k}\mathbf{k}'\mathbf{q}} d^{\dagger}_{\alpha, \sigma, \mathbf{k}+\mathbf{q}} d_{\beta, \sigma, \mathbf{k}} d^{\dagger}_{\gamma, \bar{\sigma}, \mathbf{k}'} d_{\delta, \bar{\sigma}, \mathbf{k}'+\mathbf{q}} \sum_{l} U_l [\mathbf{A}^{-1}_{\mathbf{k}+\mathbf{q}}]_{l, \alpha} [\mathbf{A}^{-1}_{\mathbf{k}}]^{*}_{l, \beta} [\mathbf{A}^{-1}_{\mathbf{k}'}]_{l, \gamma} [\mathbf{A}^{-1}_{\mathbf{k}'+\mathbf{q}}]^{*}_{l, \delta} \mathcal{F}_{\mathbf{k}\mathbf{k}'\mathbf{q}, l} \tag{29}$$

The above expression is fully general, but we remind the reader that in our case $[\mathbf{A}^{-1}_{\mathbf{k}}]$ is purely real. Also, in the case that is relevant to us, $U_l = \delta_{l, 0} U$, and the corresponding $\mathcal{F}_{\mathbf{k}\mathbf{k}'\mathbf{q}, l=0} = 1$.

We can now introduce

$$V^{\alpha\beta\gamma\delta}_{\mathbf{k}\mathbf{k}'\mathbf{q}} \equiv \sum_{l} U_l [\mathbf{A}^{-1}_{\mathbf{k}+\mathbf{q}}]_{l, \alpha} [\mathbf{A}^{-1}_{\mathbf{k}}]_{l, \beta} [\mathbf{A}^{-1}_{\mathbf{k}'}]_{l, \gamma} [\mathbf{A}^{-1}_{\mathbf{k}'+\mathbf{q}}]_{l, \delta} \mathcal{F}_{\mathbf{k}\mathbf{k}'\mathbf{q}, l} \tag{30}$$

and finally we Fourier transform to real (i.e. Wannier) space

$$\hat{H}_{\text{int}} = \frac{1}{2N} \sum_{\sigma, \alpha\beta\gamma\delta} \sum_{\mathbf{k}\mathbf{k}'\mathbf{q}} V^{\alpha\beta\gamma\delta}_{\mathbf{k}\mathbf{k}'\mathbf{q}} d^{\dagger}_{\alpha, \sigma, \mathbf{k}+\mathbf{q}} d_{\beta, \sigma, \mathbf{k}} d^{\dagger}_{\gamma, \bar{\sigma}, \mathbf{k}'} d_{\delta, \bar{\sigma}, \mathbf{k}'+\mathbf{q}} \tag{31}$$

$$= \frac{1}{2N^3} \sum_{\sigma, \alpha\beta\gamma\delta} \sum_{\mathbf{r}_1 \ldots \mathbf{r}_4} d^{\dagger}_{\alpha, \sigma, \mathbf{r}_1} d_{\beta, \sigma, \mathbf{r}_2} d^{\dagger}_{\gamma, \bar{\sigma}, \mathbf{r}_3} d_{\delta, \bar{\sigma}, \mathbf{r}_4} \sum_{\mathbf{k}\mathbf{k}'\mathbf{q}} e^{i(\mathbf{k}+\mathbf{q}) \cdot \mathbf{r}_1} e^{-i\mathbf{k} \cdot \mathbf{r}_2} e^{i\mathbf{k}' \cdot \mathbf{r}_3} e^{-i(\mathbf{k}'+\mathbf{q}) \cdot \mathbf{r}_4} V^{\alpha\beta\gamma\delta}_{\mathbf{k}\mathbf{k}'\mathbf{q}} \tag{32}$$

$$= \frac{1}{2} \sum_{\sigma, \alpha\beta\gamma\delta} \sum_{\mathbf{r}_1 \ldots \mathbf{r}_4} d^{\dagger}_{\alpha, \sigma, \mathbf{r}_1} d_{\beta, \sigma, \mathbf{r}_2} d^{\dagger}_{\gamma, \bar{\sigma}, \mathbf{r}_3} d_{\delta, \bar{\sigma}, \mathbf{r}_4} \frac{1}{N^3} \sum_{\mathbf{k}\mathbf{k}'\mathbf{q}} e^{-i\mathbf{k} \cdot (\mathbf{r}_2 - \mathbf{r}_1)} e^{-i\mathbf{k}' \cdot (\mathbf{r}_4 - \mathbf{r}_3)} e^{-i\mathbf{q} \cdot (\mathbf{r}_4 - \mathbf{r}_1)} V^{\alpha\beta\gamma\delta}_{\mathbf{k}\mathbf{k}'\mathbf{q}} \tag{33}$$

We introduce the coupling constant in real space

$$V^{\alpha\beta\gamma\delta}_{\mathbf{d}\mathbf{d}'\mathbf{u}} = \frac{1}{N^3} \sum_{\mathbf{k}\mathbf{k}'\mathbf{q}} e^{-i\mathbf{k} \cdot \mathbf{d}} e^{-i\mathbf{k}' \cdot \mathbf{d}'} e^{-i\mathbf{q} \cdot \mathbf{u}} V^{\alpha\beta\gamma\delta}_{\mathbf{k}\mathbf{k}'\mathbf{q}} \tag{34}$$

with the identification $\mathbf{d} = \mathbf{r}_2 - \mathbf{r}_1$, $\mathbf{d} = \mathbf{r}_4 - \mathbf{r}_3$, and $\mathbf{u} = \mathbf{r}_4 - \mathbf{r}_1$ and finally write

$$\hat{H}_{\mathrm{int}} = \frac{1}{2} \sum_{\sigma,\alpha\beta\gamma\delta} \sum_{\mathbf{rdd'u}} V_{\mathbf{dd'u}}^{\alpha\beta\gamma\delta} d_{\alpha,\sigma,\mathbf{r}}^{\dagger} d_{\beta,\sigma,\mathbf{r}+\mathbf{d}} d_{\gamma,\bar{\sigma},\mathbf{r}+\mathbf{u}-\mathbf{d'}}^{\dagger} d_{\delta,\bar{\sigma},\mathbf{r}+\mathbf{u}} \tag{35}$$

The interactions in Eq. 35 are all possible 4-point interactions between two electrons of opposing spins. We list below the physical meaning of some couplings:

- $V_{\mathbf{000}}^{\alpha\beta\gamma\delta}$ local coupling.

- $V_{\mathbf{000}}^{\alpha\alpha\alpha\alpha}$ local density-density coupling in the band $\alpha$.

- $V_{\mathbf{dd'u}}^{\alpha\alpha\alpha\alpha}$ intraband coupling.

- $V_{\mathbf{dd'u}}^{\alpha\alpha\beta\gamma}$ assisted interband hybridization (assisted by an operator in $\alpha$-band; hybridization is from $\gamma$ to $\beta$ band).

- $V_{\mathbf{0,0,u}}^{\alpha\alpha\beta\beta}$ density-density coupling

- $V_{\mathbf{d,0,0}}^{\alpha\alpha\beta\beta}$ density-assisted hopping in the $\alpha$ band (assisted by the density in the $\beta$ band).

- $V_{\mathbf{d,d,d}}^{\alpha\alpha\alpha\alpha}$ (intraband) pair-hopping

- $-V_{\mathbf{d,-d,0}}^{\alpha\alpha\alpha\alpha}$ $xy$-plane spin-spin interaction. This can be understood by commuting the operators $\sum_{\sigma} d_{\alpha,\sigma,\mathbf{r}}^{\dagger} d_{\alpha,\sigma,\mathbf{r}+\mathbf{d}} d_{\alpha,\bar{\sigma},\mathbf{r}+\mathbf{d}}^{\dagger} d_{\alpha,\bar{\sigma},\mathbf{r}} = -\sum_{\sigma} d_{\alpha,\sigma,\mathbf{r}}^{\dagger} d_{\alpha,\sigma,\mathbf{r}} d_{\alpha,\bar{\sigma},\mathbf{r}+\mathbf{d}}^{\dagger} d_{\alpha,\bar{\sigma},\mathbf{r}+\mathbf{d}} = -(S_{\alpha,\mathbf{r}}^{+} S_{\alpha,\mathbf{r}+\mathbf{d}}^{-} + S_{\alpha,\mathbf{r}}^{-} S_{\alpha,\mathbf{r}+\mathbf{d}}^{+}) = -2(S_{\alpha,\mathbf{r}}^{x} S_{\alpha,\mathbf{r}+\mathbf{d}}^{x} + S_{\alpha,\mathbf{r}}^{y} S_{\alpha,\mathbf{r}+\mathbf{d}}^{y})$, where $S^{x} = \frac{1}{2}(S^{+} + S^{-})$ and $S^{y} = \frac{1}{2i}(S^{+} - S^{-})$.

### C. Symmetries

The first sanity check for the coupling constant comes from the requirement that the Hamiltonian is Hermitan. The coupling constant must satisfy:

$$V_{\mathbf{dd'u}}^{\alpha\beta\gamma\delta} = V_{-\mathbf{d'},-\mathbf{d},-\mathbf{u}}^{\delta\gamma\beta\alpha} \tag{36}$$

and we check that it does.

We now restrict to $U_l = \delta_{l,0} U$. Under this assumption, with the shift $\varepsilon_d \to \varepsilon_d - U/2$, we can fully equivalently rewrite the interacting term as

$$\hat{H}_{\mathrm{int}} = \frac{1}{2} \sum_{\sigma\sigma',\alpha\beta\gamma\delta} \sum_{\mathbf{rdd'u}} V_{\mathbf{dd'u}}^{\alpha\beta\gamma\delta} d_{\alpha,\sigma,\mathbf{r}}^{\dagger} d_{\beta,\sigma,\mathbf{r}+\mathbf{d}} d_{\gamma,\sigma',\mathbf{r}+\mathbf{u}-\mathbf{d'}}^{\dagger} d_{\delta,\sigma',\mathbf{r}+\mathbf{u}} \tag{37}$$

Because of the shift in $\varepsilon_d$, the coupling constant $V_{\mathbf{dd'u}}^{\alpha\beta\gamma\delta}$ (as well as $t_{\alpha,\mathbf{d}}$) will be different. However, this rewriting of the Hamiltonian is manifestly spin-rotationally invariant. The rotation of the reference frame for the spin projection is in general performed by the following transformation

$$c_{\sigma}^{\dagger} = \sum_{\sigma'} [e^{\frac{i}{2}\theta\boldsymbol{\sigma}^{\eta}}]_{\sigma,\sigma'} c_{\sigma'}^{\dagger} = a_{\sigma\uparrow,\eta\theta} c_{\uparrow}^{\dagger} + b_{\sigma\downarrow,\eta\theta} c_{\downarrow}^{\dagger}, \quad c_{\sigma} = a_{\sigma\uparrow,\eta\theta}^{*} c_{\uparrow} + b_{\sigma\downarrow,\eta\theta}^{*} c_{\downarrow} \tag{38}$$

where $\boldsymbol{\sigma}^{\eta}$ are the Pauli matrices, $\eta = x, y, z$ is the axis of rotation, and $\theta$ is the angle of rotation. We have checked numerically that a Hamiltonian term of the form

$$\sum_{\sigma\sigma'} c_{i,\sigma}^{\dagger} c_{j,\sigma} c_{k,\sigma'}^{\dagger} c_{l,\sigma'} \tag{39}$$

is invariant under transformations of the form Eq. 38, regardless of the orbital indices $i, j, k, l$. Similar can be shown for the hopping terms of the form

$$\sum_{\sigma} c_{i,\sigma}^{\dagger} c_{j,\sigma}. \tag{40}$$

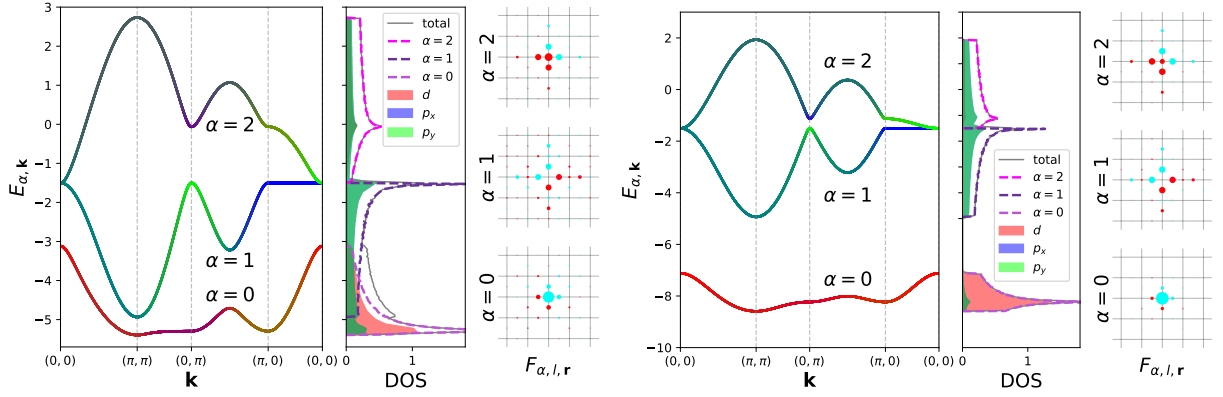

Figure 4. Illustration of the effect of the $\varepsilon_d \to \varepsilon_d - U/2$ shift on the non-interacting Hamiltonian and the effective Wannier orbitals, data point (11a). Left: Model A, no shift, corresponding to Eq. 35 formulation of the interacting part; Right: Model B, with the shift, corresponding to Eq. 37 formulation of the interacting part.

Therefore, the Hamiltonian written as Eq.37 (with the corresponding shift $\varepsilon_d \to \varepsilon_d - U/2$ in the non-interacting part) is rotationally invariant, term by term. This cannot be said for the writing in terms of Eq.35 and the original $\varepsilon_d$. The original Hamiltonian is spin-rotationally invariant (when rotation is applied to $c^\dagger_{l,\sigma,\mathbf{r}}$ and $c_{l,\sigma,\mathbf{r}}$), but, because of the non-local nature of the interaction Eq.35, when rotations are applied to $d^\dagger_{\alpha,\sigma,\mathbf{r}}$ and $d_{\alpha,\sigma,\mathbf{r}}$, this might yield same-spin interaction terms: the Hamiltonian written as Eq.35 is not spin-rotationally invariant. We refer to the non-spin-symmetric Hamiltonian as the Model A, and to the spin-symmetric Hamiltonian as the Model B.

We finally look at the spatial symmetries of the coupling constant. We check how the coupling constant $V^{\alpha\beta\gamma\delta}_{\mathbf{dd'u}}$ changes when square-lattice symmetry operations are applied to the arguments $\mathbf{dd'u}$. There are 8 symmetry operations for the square lattice, that we denote $R_m(\mathbf{r})$. The symmetry operations are $x \to -x$, $y \to -y$, $x \leftrightarrow y$ and any combination of these. We find that

$$V^{\alpha\beta\gamma\delta}_{\mathbf{dd'u}} = \pm V^{\alpha\beta\gamma\delta}_{R_m(\mathbf{d}),R_m(\mathbf{d'}),R_m(\mathbf{u})}. \tag{41}$$

The minus sign only appears in off-diagonal interactions involving the band $\alpha = 1$, more precisely the assisted hybridizations into or from the band $\alpha = 1$. It turns out that the coupling constants $V^{\alpha\beta\gamma\delta}_{\mathbf{dd'u}}$ with $(\delta_{\alpha,1} + \delta_{\beta,1} + \delta_{\gamma,1} + \delta_{\delta,1}) \bmod 2 = 1$ exhibit $d$-wave symmetry, i.e. they change sign upon $\pi/2$ rotation (around the $z$-axis) of the spatial indicies $\mathbf{d}, \mathbf{d'}, \mathbf{u}$. The change of sign upon rotation or swapping axes can be formulated as

$$V^{\alpha\beta\gamma\delta}_{\mathbf{dd'u}} = (-1)^{\delta_{\alpha,1} + \delta_{\beta,1} + \delta_{\gamma,1} + \delta_{\delta,1}} V^{\alpha\beta\gamma\delta}_{(\pm d_y, \pm d_x),(\pm d'_y, \pm d'_x),(\pm u_y, \pm u_x)}. \tag{42}$$

The absence of full square-lattice symmetry in our coupling constant is not surprising. The original model also does not have full lattice symmetry. This is already clear from the Hamiltonian Eq.3. This has an effect on correlators of the form $\rho_{ll',\mathbf{r}-\mathbf{r'}} = \langle c^\dagger_{l,\sigma,\mathbf{r}} c_{l',\sigma,\mathbf{r'}} \rangle$. Indeed, the single-particle density matrix has full square-lattice symmetry only for $l = l'$. With $l = 1$, $l' = 2$, $\rho_{ll'}$ has $d$-wave symmetry. This is reflected in the coupling constants in our rewriting of the Emery model. However, the non-interacting part (i.e. $t_{\alpha,\mathbf{d}}$) in our rewriting has full square-lattice symmetry, and so do all the intraband interactions $V^{\alpha\alpha\alpha\alpha}_{\mathbf{dd'u}}$. It is only a particular subset of the coupling constants that exhibits $d$-wave symmetry.

Finally, we observe numerically that

$$V^{\alpha\alpha\alpha\alpha}_{\mathbf{d},-\mathbf{d},\mathbf{0}} = V^{\alpha\alpha\alpha\alpha}_{\mathbf{0},\mathbf{0},\mathbf{d}} = V^{\alpha\alpha\alpha\alpha}_{\mathbf{d},\mathbf{d},\mathbf{d}}. \tag{43}$$

Some examples of spatial dependence of the interaction are given in Figures 5 and 6.

## II. RESULTS

We start from the Table 1 published in the Supplemental Material of Ref. 1 (i.e. Ref. 2) where Emery model parameters are summarized for 16 different compounds. For two 3-layer compounds, the parameters are given separately for the inner and outer layer. We list the parameters in Table I.

For the diagonalization of the non-interacting part and our rewriting of the Hamiltonian, only the difference $\varepsilon_d - \varepsilon_p$ matters, not the two parameters independently. However, the values of $\varepsilon_d$ must be additionally shifted to avoid double counting of

## data point (11a)

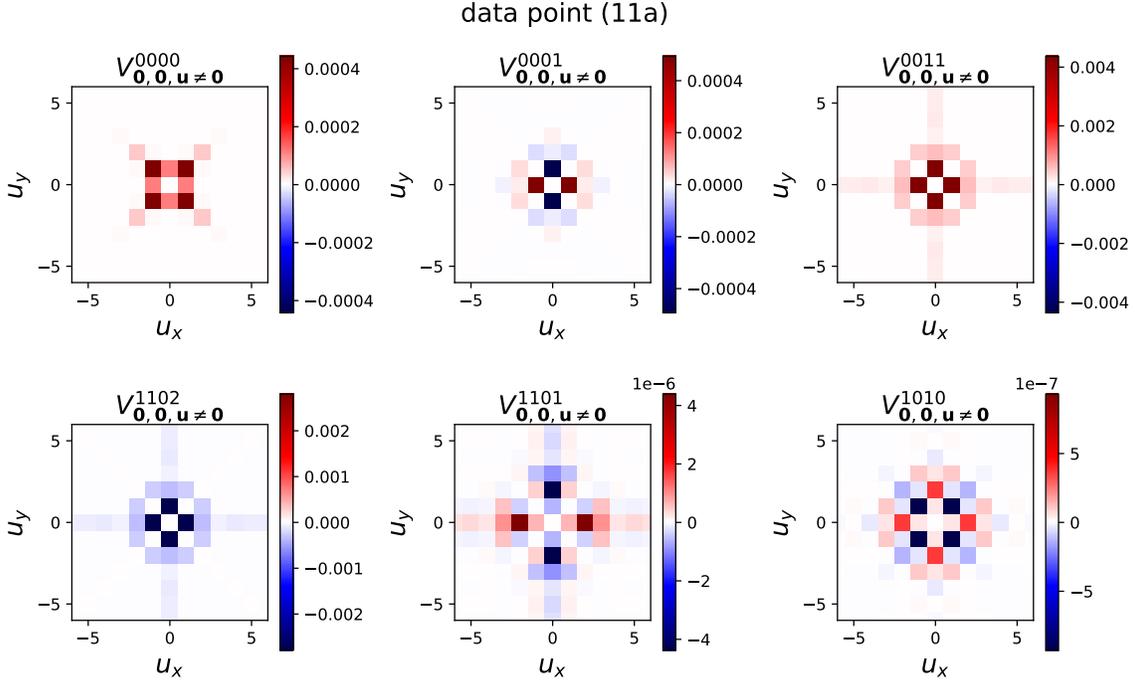

Figure 5. Illustration of the symmetry of $V_{\mathbf{dd'u}}^{\alpha\beta\gamma\delta}$, in Model A formulation. Most of the components have full square lattice-symmetry. The components representing assisted hybridization into and out of the band $\alpha = 1$ have $d$-wave symmetry. All interactions are computed on a $12 \times 12$ lattice. Some interactions have a longer range than is the size of the lattice, and in our analyses, we only focus on the short-distance components. Throughout the work, $V_{\mathbf{dd'u}}^{\alpha\beta\gamma\delta}$ is expressed in the units of $U$.

Table I. The Table with Emery model parameters for different compounds that is the basis for this work (taken from Ref.[2]).

| index | Compound | $\varepsilon_d - \varepsilon_p$ [eV] | $t_{pd}$ [eV] | $t_{pp}$ [eV] | $t'_{pp}$ [eV] | $T_c$ [K] |
|---|---|---|---|---|---|---|
| (1) | La$_2$CuO$_4$ | 2.61 | 1.39 | 0.64 | 0.103 | 38 |
| (2) | Pb$_2$Sr$_2$YCu$_3$O$_8$ | 2.32 | 1.3 | 0.673 | 0.16 | 70 |
| (3) | Ca$_2$CuO$_2$Cl$_2$ | 2.21 | 1.27 | 0.623 | 0.132 | 26 |
| (4) | La$_2$CaCu$_2$O$_6$ | 2.2 | 1.31 | 0.644 | 0.152 | 45 |
| (5) | Sr$_2$Nd$_2$NbCu$_2$O$_{10}$ | 2.1 | 1.25 | 0.612 | 0.144 | 28 |
| (6) | Bi$_2$Sr$_2$CuO$_6$ | 2.06 | 1.36 | 0.677 | 0.153 | 24 |
| (7) | YBa$_2$Cu$_3$O$_7$ | 2.05 | 1.28 | 0.673 | 0.15 | 93 |
| (8) | HgBa$_2$CaCu$_2$O$_6$ | 1.93 | 1.28 | 0.663 | 0.187 | 127 |
| (9) | HgBa$_2$CuO$_4$ | 1.93 | 1.25 | 0.649 | 0.161 | 90 |
| (10) | Sr$_2$CuO$_2$Cl$_2$ | 1.87 | 1.15 | 0.59 | 0.14 | 30 |
| (11a) | HgBa$_2$Ca$_2$Cu$_3$O$_8$(outer) | 1.87 | 1.29 | 0.674 | 0.184 | 135 |
| (11b) | HgBa$_2$Ca$_2$Cu$_3$O$_8$(inner) | 1.94 | 1.29 | 0.656 | 0.167 | 135 |
| (12) | Tl$_2$Ba$_2$CuO$_6$ | 1.79 | 1.27 | 0.63 | 0.15 | 90 |
| (13) | LaBa$_2$Cu$_3$O$_7$ | 1.77 | 1.13 | 0.62 | 0.188 | 79 |
| (14) | Bi$_2$Sr$_2$CaCu$_2$O$_8$ | 1.64 | 1.34 | 0.647 | 0.133 | 95 |
| (15) | Tl$_2$Ba$_2$CaCu$_2$O$_8$ | 1.27 | 1.29 | 0.638 | 0.14 | 110 |
| (16a) | Bi$_2$Sr$_2$Ca$_2$Cu$_3$O$_{10}$(outer) | 1.24 | 1.32 | 0.617 | 0.159 | 108 |
| (16b) | Bi$_2$Sr$_2$Ca$_2$Cu$_3$O$_{10}$(inner) | 2.24 | 1.32 | 0.678 | 0.198 | 108 |

correlation effects. We follow the prescription from Ref. [1], where a fixed double-counting shift $E_{\mathrm{dc}} = 3.12$ eV was subtracted from $\varepsilon_d$, independently of compound (and, in their calculations, independently of doping).

Based on the model parameters from the Table I, we compute the following, for each of the compounds:

- the total bandwidth $D = \max_{\mathbf{k},\alpha} E_{\alpha,\mathbf{k}} - \min_{\mathbf{k},\alpha} E_{\alpha,\mathbf{k}}$

- the bandwidths of each band $D_\alpha = \max_{\mathbf{k}} E_{\alpha,\mathbf{k}} - \min_{\mathbf{k}} E_{\alpha,\mathbf{k}}$

- the overlap/gap between the bottom two bands $O = \max_{\mathbf{k}} E_{\alpha=0,\mathbf{k}} - \min_{\mathbf{k}} E_{\alpha=1,\mathbf{k}}$

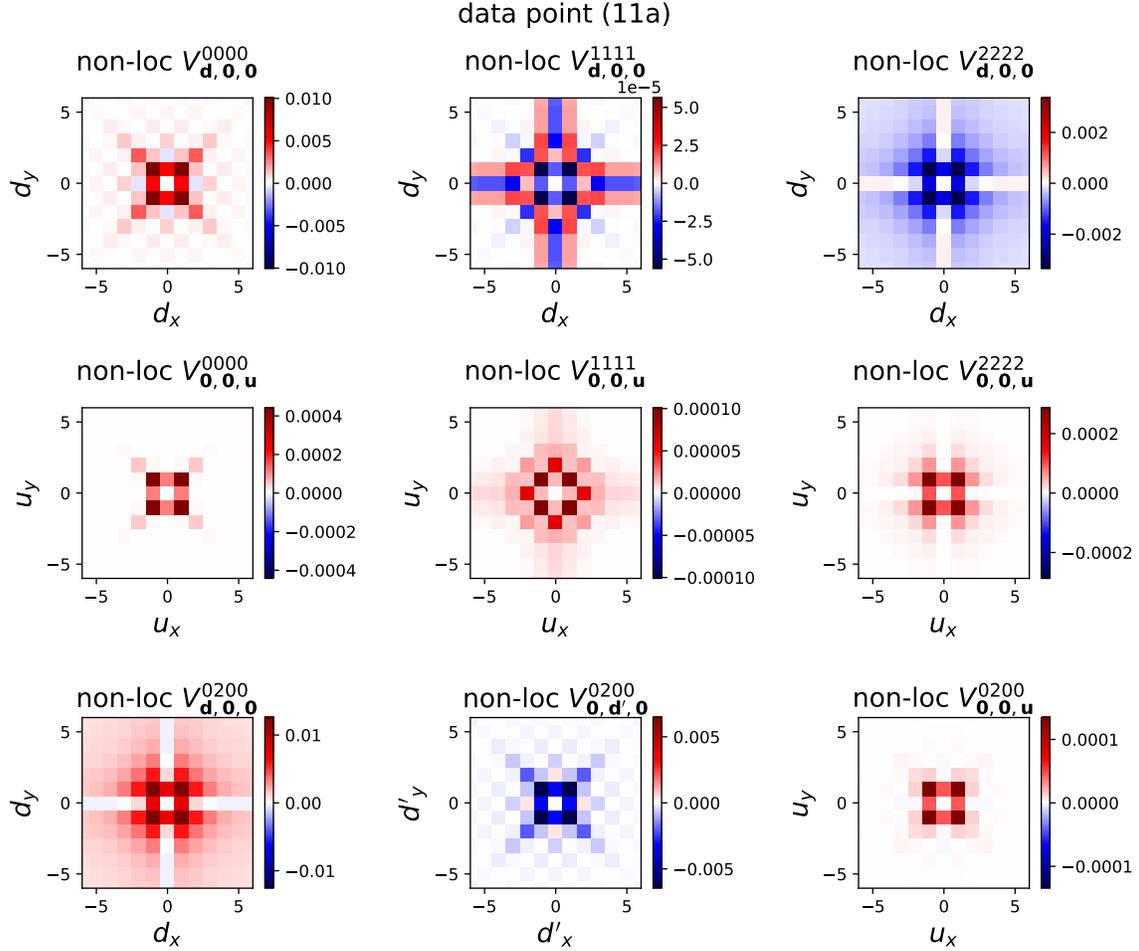

Figure 6. Illustration of the spatial dependence of relevant interaction components (Model A formulation).

- the gap between $\alpha = 0$ and $\alpha = 1$ bands at $\mathbf{k}$-points $(0,0)$ and $(\pi, \pi)$

- the gap between $\alpha = 1$ and $\alpha = 2$ bands at $\mathbf{k} = (0, \pi)$

- the short-distance hoppings $t_\alpha \equiv t_{\alpha, \mathbf{d}=(1,0)}$, $t'_\alpha \equiv t_{\alpha, \mathbf{d}=(1,1)}$, $t''_\alpha \equiv t_{\alpha, \mathbf{d}=(2,0)}$

- the ratios $t'_\alpha / t_\alpha$ and $t''_\alpha / t_\alpha$

- the total amount of $d$-character in the $\alpha = 2$ band, $\frac{1}{N} \sum_{\mathbf{k}} |[\mathbf{A_k}]_{\alpha=2, l=0}|^2$

- the total amount of $p$-character in the $\alpha = 0$ band, $\frac{1}{N} \sum_{l=1,2} \sum_{\mathbf{k}} |[\mathbf{A_k}]_{\alpha=0, l}|^2$

- the maximum and the minimum value in $V^{\alpha\beta\gamma\delta}_{\mathbf{dd'u}}$

- the minimum intraband coupling constant for each band, $\min_{\mathbf{dd'u}} V^{\alpha\alpha\alpha\alpha}_{\mathbf{dd'u}}$

- the local density-density interaction in each band $V^{\alpha\alpha\alpha\alpha}_{\mathbf{000}}$

- the nearest ($\mathbf{d} = (1,0)$) and the next-nearest neighbor ($\mathbf{d} = (1,1)$) assisted hopping $V^{\alpha\alpha\alpha\alpha}_{\mathbf{d,00}}$ and the spin-spin interaction $V^{\alpha\alpha\alpha\alpha}_{\mathbf{d,-d,0}}$ in each of the bands

Throughout the paper, we give $V^{\alpha\beta\gamma\delta}_{\mathbf{dd'u}}$ in units of $U$ (the estimate is $U = 8$ eV for all compounds), while the rest of the parameters are given in units of eV.

After computing the above parameters, we do a linear fit of $T_c$ to each of the parameters individually, and compute the std. deviation. The results are summarized in Table II. We compare the results for the non-spin-symmetric (Model A, corresponding

to Eq. 35; see Table II.A) and spin-symmetric Hamiltonians (Model B, corresponding to Eq. 37 and a shift $\varepsilon_d \to \varepsilon_d - U/2$; see Table II.B). We also show the difference in results depending on whether points (1),(6) and (11b) are excluded or not (Table II.A.1 vs. Table II.A.2, and Table II.B.1 vs. Table II.B.2).

Table II. Top list of best single-parameter predictors of $T_c$.

Table II.A. Model A (non spin-rotationally symmetric Hamiltonian).

Table II.A.1. All data points included.

| rank | parameter | std. dev. $\sigma$ |
|---|---|---|
| 1 | $t'/t$ | 27.4927 |
| 2 | $t'_{\alpha=2}$ | 27.5412 |
| 3 | $t_{\alpha=1}$ | 29.5241 |
| 4 | $V^{0000}_{(1,0),(-1,0),\mathbf{0}}$ | 29.6248 |
| 5 | $t''_{\alpha=1}$ | 29.9415 |
| 6 | $D_{\alpha=1}$ | 30.2950 |
| 7 | $t'_{pp}$ | 30.5228 |
| 8 | $V^{0000}_{(1,0),\mathbf{0},\mathbf{0}}$ | 30.8614 |
| 9 | $t'_{\alpha=2}/t_{\alpha=2}$ | 31.4597 |
| 10 | $E^{\alpha=1}_{\mathbf{k}=(0,0)} - E^{\alpha=0}_{\mathbf{k}=(0,0)}$ | 31.7394 |

Table II.A.2. Points (1),(6) and (11b) excluded.

| rank | parameter | std. dev. $\sigma$ |
|---|---|---|
| 1 | $t'_{\alpha=2}$ | 21.7034 |
| 2 | $t''_{\alpha=2}$ | 23.3602 |
| 3 | $t_{\alpha=1}$ | 23.6241 |
| 4 | $t''_{\alpha=1}$ | 24.0605 |
| 5 | $D_{\alpha=1}$ | 24.8676 |
| 6 | $t'/t$ | 24.9718 |
| 7 | $V^{0000}_{(1,0),(-1,0),\mathbf{0}}$ | 26.9649 |
| 8 | $t_{pp}$ | 27.0259 |
| 9 | $V^{0000}_{(1,0),\mathbf{0},\mathbf{0}}$ | 27.5327 |
| 10 | $t'_{pp}$ | 28.1611 |

Table II.B. Model B (spin-rotationally symmetric Hamiltonian).

Table II.B.1. All data points included.

| rank | parameter | std. dev. $\sigma$ |
|---|---|---|
| 1 | $t'_{\alpha=2}$ | 26.1661 |
| 2 | $t'/t$ | 27.4927 |
| 3 | $D_{\alpha=1}$ | 30.2950 |
| 4 | $t_{\alpha=1}$ | 30.4133 |
| 5 | $t'_{pp}$ | 30.5228 |
| 6 | $E^{\alpha=1}_{\mathbf{k}=(0,0)} - E^{\alpha=0}_{\mathbf{k}=(0,0)}$ | 31.7394 |
| 7 | $t''_{\alpha=2}/t_{\alpha=2}$ | 32.8284 |
| 8 | $t''_{\alpha=1}$ | 33.0235 |
| 9 | $t''_{\alpha=0}$ | 33.0360 |
| 10 | $\varepsilon_d - \varepsilon_p$ | 33.1773 |

Table II.B.2. Points (1),(6) and (11b) excluded.

| rank | parameter | std. dev. $\sigma$ |
|---|---|---|
| 1 | $t'_{\alpha=2}$ | 16.7233 |
| 2 | $D_{\alpha=1}$ | 24.8676 |
| 3 | $t_{\alpha=1}$ | 24.8911 |
| 4 | $t'/t$ | 24.9718 |
| 5 | $D$ | 26.6340 |
| 6 | $t''_{\alpha=2}$ | 26.7989 |
| 7 | $t''_{\alpha=1}$ | 26.9456 |
| 8 | $t_{pp}$ | 27.0259 |
| 9 | $t'_{pp}$ | 28.1611 |
| 10 | $t'_{\alpha=1}$ | 28.2256 |

Next, we look at all possible linear combinations of two parameters, $P(p_1, p_2) = c_1 p_1 + c_2 p_2$, with $c_1^2 + c_2^2 = 1$. For each pair $(p_1, p_2)$ we minimize the std. deviation $\sigma$ w.r.t. $c_1$ and $c_2$, and sort the pairs according to the minimal $\sigma$. The results are given in Table III.

Finally, we look at all possible linear combinations of three parameters, $P(p_1, p_2, p_3) = c_1 p_1 + c_2 p_2 + c_3 p_3$, with $c_1^2 + c_2^2 + c_3^2 = 1$, and repeat the optimization procedure. The top lists of best 3-parameter predictors, based on the models A and B, are given in Table IV. We note that there are about 30 3-parameter predictors based on model B parameters that are better than the best 3-parameter predictor based on model A parameters. Finally, we identify the parameters which appear the most in the linear combinations which constitute the top 100 predictors. The results for model A and B, separately, are given in the Table V.

The results are illustrated in Fig. 7. We note that the two best 3-parameter predictors (in both Model A and Model B) differ by only one parameter, and yield a very similar $\sigma$. The parameters that differ, therefore, must correlate well, which we check in the bottom row.

We now restrict to the original parameters of the Emery model, i.e. $\varepsilon_d - \varepsilon_p, t_{pd}, t_{pp}$ and $t'_{pp}$, and the $t'/t$ discussed in *Weber et al.*. We show corresponding fits in Fig. 8. The two best 2-parameter predictors and the best 3-parameter predictor are given in the bottom row. In the panel $g$, we show how the std. deviation depends on the choice of coefficients $c_{1..3}$, for the linear combination of $\varepsilon_d - \varepsilon_p, t_{pp}$ and $t'_{pp}$, which yields the best fit of $T_c$. The collapse of the data points that we can obtain this way is noticeably worse than what we can do with the parameters of the models A or B.

In the end, we want to estimate the probability that the scaling with $T_c$ we get for some parameters and their linear combinations is due to pure chance. We repeat the entire procedure with all parameters and variables (except $T_c$) replaced by random numbers between 0 and 1. We also exclude the points (1),(6) and (11b). We get the following results:

- best single-parameter predictor: $\sigma \approx 27K$

- best two-parameter linear-combination: $\sigma \approx 19K \approx 27/\sqrt{2}$

- best three-parameter linear-combination: $\sigma \approx 15K \approx 27/\sqrt{3}$

Table III. Best linear combinations of two parameters.

Table III.A. Model A.

| rank | $p_1$ | $p_2$ | $c_1$ | $c_2$ | std. dev. $\sigma$ |
|---|---|---|---|---|---|
| 1 | $\min V_{\mathbf{d},\mathbf{d}',\mathbf{u}}^{\alpha\beta\gamma\delta} = V_{\mathbf{0},\mathbf{0},\mathbf{0}}^{0020}$ | $t''_{\alpha=1}$ | -0.422894 | 0.906179 | 5.4231 |
| 2 | $\min V_{\mathbf{d},\mathbf{d}',\mathbf{u}}^{\alpha\beta\gamma\delta} = V_{\mathbf{0},\mathbf{0},\mathbf{0}}^{0020}$ | $t_{\alpha=1}$ | -0.894717 | 0.446634 | 5.7230 |
| 3 | $D_{\alpha=1}$ | $\min V_{\mathbf{d},\mathbf{d}',\mathbf{u}}^{\alpha\beta\gamma\delta} = V_{\mathbf{0},\mathbf{0},\mathbf{0}}^{0020}$ | 0.060565 | -0.998164 | 5.7304 |
| 4 | $V_{\mathbf{0},\mathbf{0},\mathbf{0}}^{2222}$ | $t''_{\alpha=2}$ | -0.178606 | -0.983921 | 6.1861 |
| 5 | $t_{pp}$ | $V_{(1,1),\mathbf{0},\mathbf{0}}^{2222}$ | 0.012569 | -0.999921 | 6.2400 |
| 6 | $V_{\mathbf{0},\mathbf{0},\mathbf{0}}^{2222}$ | $t'_{\alpha=0}$ | -0.668412 | 0.743791 | 6.5120 |
| 7 | $\frac{1}{N}\sum_{\mathbf{k}}|[A_{\mathbf{k}}]_{\alpha=2,l=0}|^2$ | $t''_{\alpha=2}$ | -0.101612 | -0.994824 | 6.8141 |
| 8 | $\frac{1}{N}\sum_{\mathbf{k}}|[A_{\mathbf{k}}]_{\alpha=2,l=0}|^2$ | $t'_{\alpha=0}$ | -0.452919 | 0.891552 | 7.2931 |
| 9 | $\min V_{\mathbf{d},\mathbf{d}',\mathbf{u}}^{\alpha\beta\gamma\delta} = V_{\mathbf{0},\mathbf{0},\mathbf{0}}^{0020}$ | $t''_{\alpha=2}$ | -0.774502 | -0.632572 | 7.4161 |
| 10 | $t'_{\alpha=1}/t_{\alpha=1}$ | $t''_{\alpha=2}$ | 0.346422 | -0.938079 | 7.4191 |

Table III.B. Model B.

| rank | $p_1$ | $p_2$ | $c_1$ | $c_2$ | std. dev. $\sigma$ |
|---|---|---|---|---|---|
| 1 | $V_{(1,1),(-1,-1),\mathbf{0}}^{0000}$ | $t'_{\alpha=1}$ | -0.999998 | -0.001861 | 5.7407 |
| 2 | $E_{\mathbf{k}=(\pi,\pi)}^{\alpha=1} - E_{\mathbf{k}=(\pi,\pi)}^{\alpha=0}$ | $t'_{\alpha=1}$ | 0.006267 | 0.999980 | 6.2854 |
| 3 | $V_{(1,1),\mathbf{0},\mathbf{0}}^{0000}$ | $t'_{\alpha=1}$ | -0.997331 | -0.073015 | 6.4198 |
| 4 | $V_{(1,1),(-1,-1),\mathbf{0}}^{0000}$ | $t''_{\alpha=2}$ | -0.999992 | -0.004007 | 6.4207 |
| 5 | $V_{(1,1),(-1,-1),\mathbf{0}}^{0000}$ | $t''_{\alpha=1}$ | -0.999994 | 0.003484 | 6.8968 |
| 6 | $V_{(1,1),\mathbf{0},\mathbf{0}}^{0000}$ | $t''_{\alpha=2}$ | -0.987743 | -0.156088 | 6.8977 |
| 7 | $V_{(1,1),\mathbf{0},\mathbf{0}}^{0000}$ | $t''_{\alpha=1}$ | -0.990743 | 0.135753 | 7.3100 |
| 8 | $\varepsilon_d - \varepsilon_p$ | $\max E_{\mathbf{k}}^{\alpha=0} - \min E_{\mathbf{k}}^{\alpha=1}$ | -0.799401 | 0.600798 | 7.5377 |
| 9 | $\varepsilon_d - \varepsilon_p$ | $t'_{\alpha=2}$ | -0.007130 | 0.999975 | 8.0288 |
| 10 | $\varepsilon_d - \varepsilon_p$ | $t_{\alpha=1}$ | -0.054257 | 0.998527 | 8.1007 |

Table IV. Top lists of 3-parameter predictors.

Table IV.A. Model A.

| rank | $p_1$ | $p_2$ | $p_3$ | $c_1$ | $c_2$ | $c_3$ | std. dev. $\sigma$ |
|---|---|---|---|---|---|---|---|
| 1 | $D_{\alpha=1}$ | $t''_{\alpha=1}$ | $t'_{\alpha=2}$ | -0.027127 | 0.997861 | 0.059480 | 4.4696 |
| 2 | $D_{\alpha=1}$ | $V_{\mathbf{0},\mathbf{0},\mathbf{0}}^{2222}$ | $t''_{\alpha=1}$ | -0.024756 | -0.034964 | 0.999082 | 4.5503 |
| 3 | $D_{\alpha=2}$ | $t'_{\alpha=0}$ | $t''_{\alpha=0}$ | -0.103772 | 0.355220 | -0.929005 | 4.5828 |
| 4 | $D_{\alpha=1}$ | $t''_{\alpha=2}$ | $t''_{\alpha=2}/t_{\alpha=2}$ | -0.014299 | -0.942122 | -0.334964 | 4.6311 |
| 5 | $D_{\alpha=1}$ | $t'_{\alpha=0}$ | $t''_{\alpha=1}$ | -0.024244 | -0.043401 | 0.998764 | 4.6446 |
| 6 | $V_{\mathbf{0},\mathbf{0},\mathbf{0}}^{2222}$ | $t_{\alpha=1}$ | $t''_{\alpha=1}/t_{\alpha=1}$ | -0.070622 | 0.051866 | 0.996154 | 4.6573 |
| 7 | $V_{\mathbf{0},\mathbf{0},\mathbf{0}}^{2222}$ | $t''_{\alpha=2}$ | $t''_{\alpha=1}/t_{\alpha=1}$ | -0.076054 | 0.243175 | 0.966996 | 4.6784 |
| 8 | $\min_{\mathbf{d},\mathbf{d}',\mathbf{u}} V_{\mathbf{d},\mathbf{d}',\mathbf{u}}^{2222} = V_{(1,1),\mathbf{0},\mathbf{0}}^{2222}$ | $t''_{\alpha=0}/t_{\alpha=0}$ | $t'_{\alpha=1}$ | -0.996019 | 0.031442 | 0.083417 | 4.7134 |
| 9 | $V_{\mathbf{0},\mathbf{0},\mathbf{0}}^{2222}$ | $\min_{\mathbf{d},\mathbf{d}',\mathbf{u}} V_{\mathbf{d},\mathbf{d}',\mathbf{u}}^{0000}$ | $t'_{\alpha=1}$ | -0.021006 | 0.999127 | -0.036103 | 4.7267 |
| 10 | $V_{\mathbf{0},\mathbf{0},\mathbf{0}}^{2222}$ | $t'_{\alpha=1}$ | $t'_{\alpha=2}/t_{\alpha=2}$ | -0.335834 | 0.930401 | 0.146868 | 4.7306 |

Table IV.B. Model B.

| rank | $p_1$ | $p_2$ | $p_3$ | $c_1$ | $c_2$ | $c_3$ | std. dev. $\sigma$ |
|---|---|---|---|---|---|---|---|
| 1 | $V_{(1,1),\mathbf{0},\mathbf{0}}^{2222}$ | $t'_{\alpha=0}$ | $t''_{\alpha=0}/t_{\alpha=0}$ | 0.999856 | 0.016821 | 0.002042 | 4.0065 |
| 2 | $\min_{\mathbf{d},\mathbf{d}',\mathbf{u}} V_{\mathbf{d},\mathbf{d}',\mathbf{u}}^{0000}$ | $V_{(1,1),\mathbf{0},\mathbf{0}}^{2222}$ | $t'_{\alpha=0}$ | 0.115210 | 0.993193 | 0.017135 | 4.0265 |
| 3 | $V_{(1,0),\mathbf{0},\mathbf{0}}^{1111}$ | $V_{(1,1),\mathbf{0},\mathbf{0}}^{2222}$ | $t'_{\alpha=0}$ | 0.999946 | 0.010371 | 0.000194 | 4.0770 |
| 4 | $V_{(1,1),\mathbf{0},\mathbf{0}}^{2222}$ | $V_{(1,1),(-1,-1),\mathbf{0}}^{0000}$ | $t'_{\alpha=0}$ | 0.313155 | -0.949673 | 0.007395 | 4.0836 |
| 5 | $V_{(1,1),(-1,-1),\mathbf{0}}^{2222}$ | $t'_{\alpha=0}$ | $t''_{\alpha=2}/t_{\alpha=2}$ | -0.999999 | 0.001642 | -0.000130 | 4.0942 |
| 6 | $E_{\mathbf{k}=(0,\pi)}^{\alpha=2} - E_{\mathbf{k}=(0,\pi)}^{\alpha=1}$ | $V_{(1,1),(-1,-1),\mathbf{0}}^{2222}$ | $t'_{\alpha=0}$ | 0.000010 | -0.999999 | 0.001553 | 4.1090 |
| 7 | $t_{pd}$ | $V_{(1,1),\mathbf{0},\mathbf{0}}^{0000}$ | $t'_{\alpha=0}$ | 0.004709 | -0.368569 | 0.929589 | 4.1152 |
| 8 | $D_{\alpha=2}$ | $V_{(1,1),(-1,-1),\mathbf{0}}^{2222}$ | $t'_{\alpha=0}$ | 0.000009 | -0.999999 | 0.001444 | 4.1644 |
| 9 | $V_{\mathbf{0},\mathbf{0},\mathbf{0}}^{2222}$ | $t_{\alpha=0}$ | $t'_{\alpha=0}/t_{\alpha=0}$ | -0.994567 | 0.093453 | 0.045869 | 4.1822 |
| 10 | $V_{(1,1),(-1,-1),\mathbf{0}}^{2222}$ | $t'_{\alpha=0}$ | $t_{\alpha=2}$ | -0.999999 | 0.001434 | -0.000071 | 4.2076 |

Comparing to our best results ($\sigma = 16K$, $\sigma = 5.4K$ and $\sigma = 4.0K$ in the three categories, respectively), we conclude that our results are highly unlikely to be obtained by pure chance, and that the quality of scaling we obtain must be indicative of actual

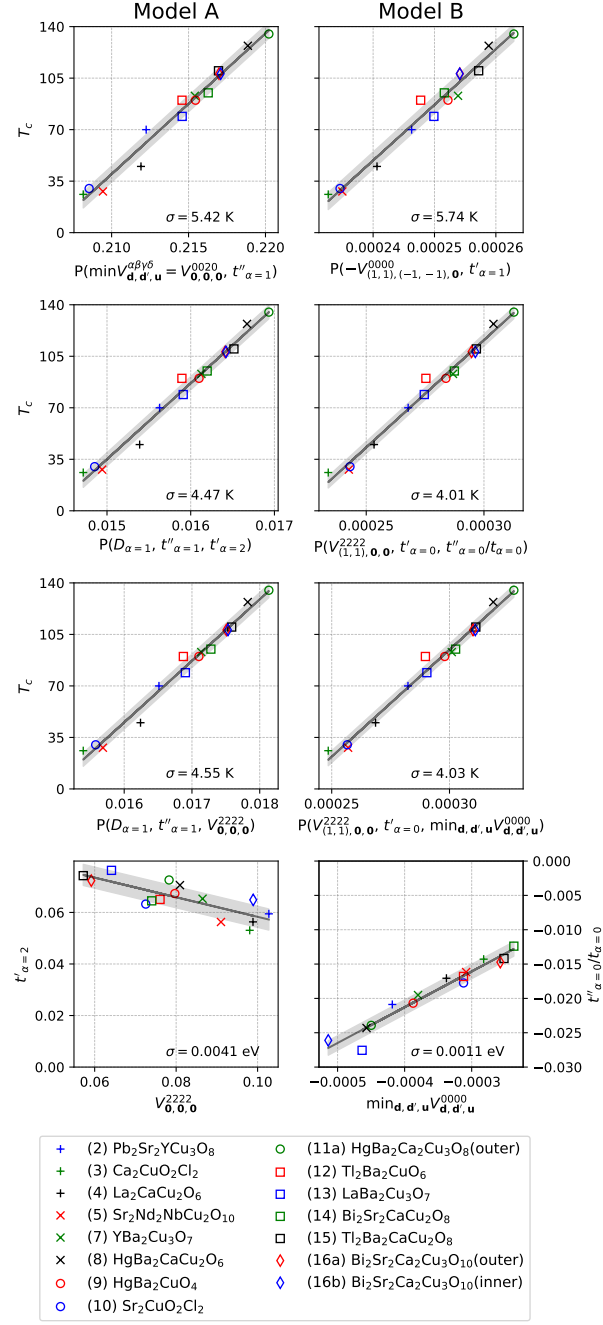

Figure 7. Top row: best 2-parameter predictors. Middle two rows: best and second best 3-parameter predictors. In both cases, only one parameter is different between the best and the second best. We check in the bottom row that these parameters correlate well. The best single parameter predictor $t'_{\alpha=2}$ correlates closely with $V_{000}^{2222}$, which, on the other hand, appears most times in the top 100 best 3-parameter predictors based on Model A. Note that $V_{\mathbf{d}\mathbf{d}'\mathbf{u}}^{\alpha\beta\gamma\delta}$ is throughout the paper given in units of $U$, while the rest of the parameters are given in units of eV.

correlations between $T_c$ and variables that we study.

## III. SCAN OF THE PARAMETER SPACE - IDENTIFYING HIGH-$T_c$ CANDIDATES

In this section we scan the parameter-space of the Emery model. For each parameter we cover the range of values spanned by the data points (1)-(16b), expanded by 33% in both directions. We discretize each parameter by 6 equidistant points, which gives

Table V. Top list of parameters, ranked by the number of times they appear in the top 100 3-parameter predictors ($N_{\text{appear.}}$).

<table>
<tr><td colspan="3" align="center">Table V.A. Model A.</td><td></td><td colspan="3" align="center">Table V.B. Model B.</td></tr>
<tr><td>rank</td><td>parameter</td><td>$N_{\text{appear.}}$</td><td></td><td>rank</td><td>parameter</td><td>$N_{\text{appear.}}$</td></tr>
<tr><td>1</td><td>$V^{2222}_{\mathbf{0,0,0}}$</td><td>40</td><td></td><td>1</td><td>$t'_{\alpha=0}$</td><td>37</td></tr>
<tr><td>2</td><td>$t'_{\alpha=0}$</td><td>30</td><td></td><td>2</td><td>$V^{2222}_{(1,1),\mathbf{0,0}}$</td><td>28</td></tr>
<tr><td>3</td><td>$t''_{\alpha=1}$</td><td>23</td><td></td><td>3</td><td>$V^{2222}_{(1,1),(-1,-1),\mathbf{0}}$</td><td>26</td></tr>
<tr><td>4</td><td>$t'_{\alpha=1}$</td><td>21</td><td></td><td>4</td><td>$V^{0000}_{(1,1),(-1,-1),\mathbf{0}}$</td><td>25</td></tr>
<tr><td>5</td><td>$t''_{\alpha=1}/t_{\alpha=1}$</td><td>17</td><td></td><td>5</td><td>$t''_{\alpha=1}$</td><td>19</td></tr>
<tr><td>6</td><td>$t_{\alpha=1}$</td><td>15</td><td></td><td>6</td><td>$V^{2222}_{\mathbf{0,0,0}}$</td><td>18</td></tr>
<tr><td>7</td><td>$\min V^{\alpha\beta\gamma\delta}_{\mathbf{d,d',u}}=V^{0020}_{\mathbf{0,0,0}}$</td><td>15</td><td></td><td>7</td><td>$D_{\alpha=0}$</td><td>15</td></tr>
<tr><td>8</td><td>$D_{\alpha=1}$</td><td>15</td><td></td><td>8</td><td>$t_{pd}$</td><td>14</td></tr>
<tr><td>9</td><td>$\min_{\mathbf{d,d',u}}V^{0000}_{\mathbf{d,d',u}}$</td><td>12</td><td></td><td>9</td><td>$t''_{\alpha=2}$</td><td>11</td></tr>
<tr><td>10</td><td>$t''_{\alpha=2}$</td><td>11</td><td></td><td>10</td><td>$t_{\alpha=0}$</td><td>11</td></tr>
</table>

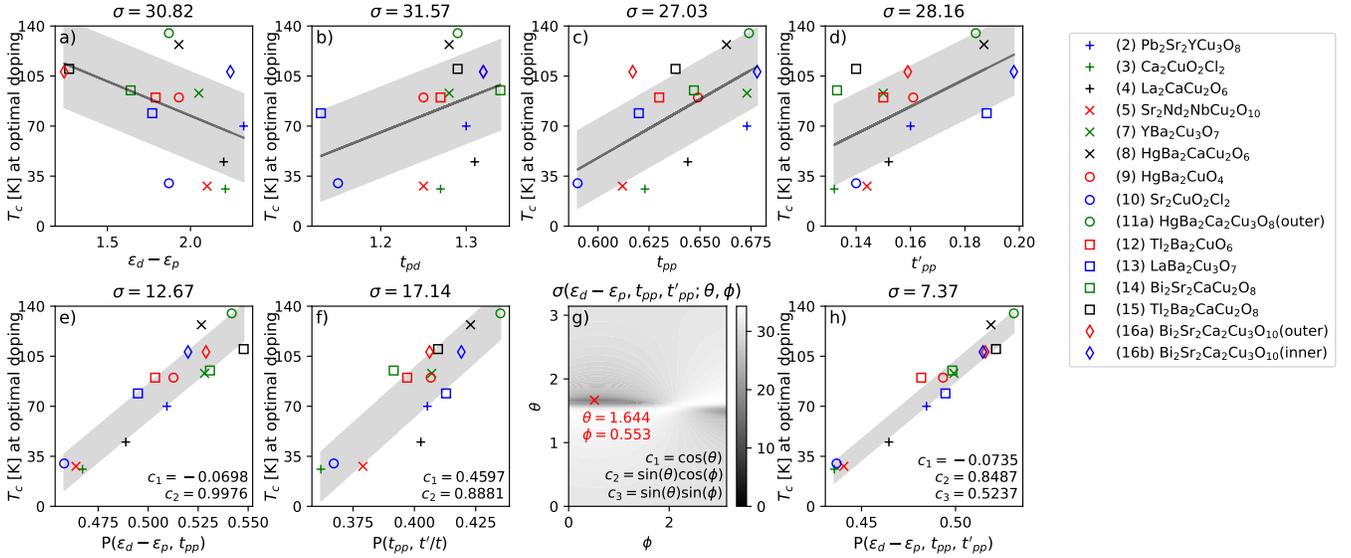

Figure 8. a)-d) The fit of $T_c$ to original parameters of the model. e)-f) the two best 2-parameter predictors based on the original parameters and $t'/t$ from *Weber et al.*. g)-h) The best 3-parameter fit based on the original parameters and $t'/t$. g) shows how the quality of the fit depends on the coefficients of the linear combination.

us in total $6^4 = 1296$ points to work with. Then, for each point we evaluate the $T_c$-prediction, given by the top 100 3-parameter predictors based on parameters of model B (the 100th best has $\sigma = 5.02$ K). We then compute the average $T_c$-prediction, $\langle T_c \rangle = \sum_P T_c(P)/N_P$, and the std. deviation, $\sigma = \sqrt{\sum_P (T_c(P) - \langle T_c \rangle)^2/N_P}$, where different predictors are denoted $P$, and $N_P = 100$.

On Fig. 9 we plot $\langle T_c \rangle$ vs. $\sigma$. We see that, in general, different predictors are correlated in a large portion of the parameter space that we study. There is plenty of cases where different predictors agree to within $\pm 10$ K. We also see that there are points inside the 4D box spanned by the *Weber et al.* dataset, where $T_c$ is predicted to be very large. In particular, if we take the minimal $\varepsilon_d - \varepsilon_p$ and the maximal $t_{pd}$, $t_{pp}$ and $t'_{pp}$ from the *Weber et al.* dataset, the $T_c$ is predicted to be 194.9±4.5 K. This is in sharp contrast with the analysis presented in *Weber et al.*, where one could only conclude that $T_c$ for this choice of parameters would be around 105±30 K. In the Table VI we list 10 high-$T_c$ candidates, where $T_c > 150$ K, and all parameters are inside the range spanned by the data points in the *Weber et al.* dataset.

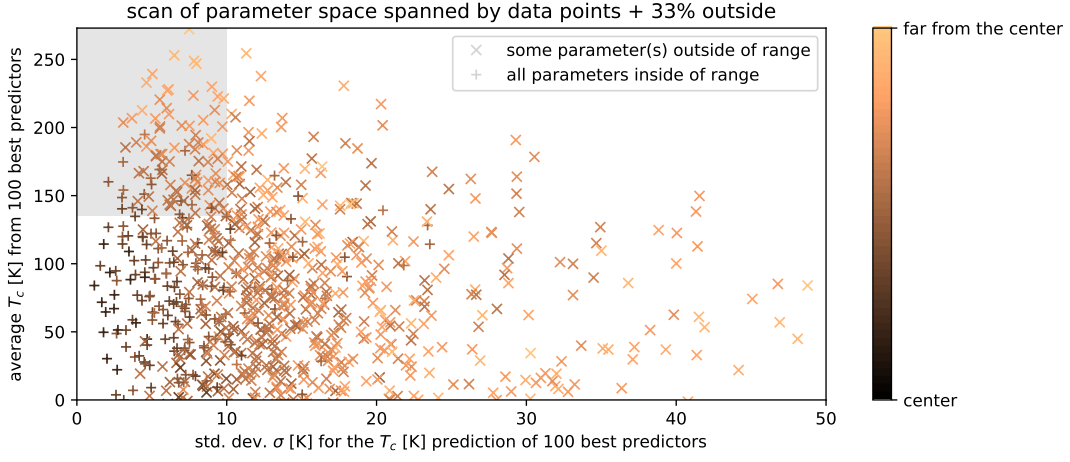

Figure 9. The scan of the parameter space spanned by the data points, enlarged by 33% in both directions, for each parameter. On the plot we show the average $T_c$ as predicted by the top 100 3-parameter predictors based on Model B, vs. the corresponding standard deviation. The gray rectangle denotes the window of opportunity: where different predictors agree well and the predicted $T_c$ is high.

Table VI. List of high-$T_c$ candidates in the parameter space spanned by the data points in the *Weber et al.* dataset. The $T_c$ estimate $\langle T_c \rangle$ is obtained by averaging predictions of the top 100 3-parameter predictors based on the parameters of the (spin-symmetric) model B, and $\sigma$ is the corresponding std. deviation. Note that the convention for $\varepsilon_d - \varepsilon_p$ is the same as in the original table in *Weber et al.*, i.e. we do not include here the $E_{dc}$ shift of $\varepsilon_d$ that is, nevertheless, used in our calculation.

| $\varepsilon_d - \varepsilon_p$ [eV] | $t_{pd}$ [eV] | $t_{pp}$ [eV] | $t'_{pp}$ [eV] | $\langle T_c \rangle$ [K] | $\sigma$ [K] |
|---|---|---|---|---|---|
| 1.240 | 1.390 | 0.678 | 0.135 | 154.1 | 3.1 |
| 1.240 | 1.390 | 0.678 | 0.166 | 174.7 | 3.1 |
| 1.240 | 1.390 | 0.678 | 0.198 | 194.9 | 4.5 |
| 1.697 | 1.217 | 0.678 | 0.198 | 151.1 | 4.5 |
| 1.240 | 1.303 | 0.649 | 0.198 | 158.2 | 5.0 |
| 1.697 | 1.390 | 0.678 | 0.198 | 162.8 | 4.9 |
| 1.240 | 1.303 | 0.678 | 0.166 | 165.0 | 7.7 |
| 1.240 | 1.303 | 0.678 | 0.198 | 182.9 | 7.2 |
| 1.697 | 1.303 | 0.678 | 0.198 | 160.2 | 2.1 |
| 1.240 | 1.390 | 0.649 | 0.198 | 169.2 | 5.5 |

## IV. RELATIONSHIP TO SUPERCONDUCTIVITY IN THE ORIGINAL MODEL

Let us assume a $d$-wave superconductivity living on the copper $d$-orbitals, and let's assume that the instantaneous anomalous correlator has the following form

$$\langle c^\dagger_{l,\uparrow,\mathbf{r}} c^\dagger_{l',\downarrow,\mathbf{r}'} \rangle = \Delta \delta_{l,0} \delta_{l',0} (\delta_{\mathbf{r}-\mathbf{r}',\mathbf{e}_x} + \delta_{\mathbf{r}-\mathbf{r}',-\mathbf{e}_x} - \delta_{\mathbf{r}-\mathbf{r}',\mathbf{e}_y} - \delta_{\mathbf{r}-\mathbf{r}',-\mathbf{e}_y}) \tag{44}$$

In terms of our Wannier orbitals, the anomalous propagator is then given by

$$\langle d^\dagger_{\alpha,\uparrow,\mathbf{r}} d^\dagger_{\beta,\downarrow,\mathbf{r}'} \rangle = \sum_{l\mathbf{r}'',l'\mathbf{r}'''} F_{\alpha,l,\mathbf{r}-\mathbf{r}''} F_{\beta,l',\mathbf{r}'-\mathbf{r}'''} \langle c^\dagger_{l,\uparrow,\mathbf{r}''} c^\dagger_{l',\downarrow,\mathbf{r}'''} \rangle \tag{45}$$

$$= \sum_{l\mathbf{r}'',l'\mathbf{d}\in\{\pm\mathbf{e}_x,\pm\mathbf{e}_y\}} F_{\alpha,l,\mathbf{r}-\mathbf{r}''} F_{\beta,l',\mathbf{r}'-\mathbf{r}''-\mathbf{d}} \langle c^\dagger_{l,\uparrow,\mathbf{r}''} c^\dagger_{l',\downarrow,\mathbf{r}''+\mathbf{d}} \rangle \tag{46}$$

$$= \sum_{\mathbf{r}'',\mathbf{d}\in\{\pm\mathbf{e}_x,\pm\mathbf{e}_y\}} F_{\alpha,0,\mathbf{r}-\mathbf{r}''} F_{\beta,0,\mathbf{r}'-\mathbf{r}''-\mathbf{d}} (-1)^{\delta_{\mathbf{d},\mathbf{e}_y}+\delta_{\mathbf{d},-\mathbf{e}_y}} \Delta \tag{47}$$

We compute $\langle d^\dagger_{\alpha,\uparrow,\mathbf{r}} d^\dagger_{\beta,\downarrow,\mathbf{0}} \rangle$ (Figure 10) and see that the dominant values are for $\alpha,\beta = 0,2$, $\mathbf{r} = (0,1)$, and they have $d$-wave symmetry. There is also a weak $s$-wave component in the pairing between $\alpha = 1$ and $\alpha = 0,2$ bands. Overall, the superconductivity of $d$-orbitals is not very different from the superconductivity in the $\alpha = 0$ and $\alpha = 2$ orbitals. This provides

some understanding of why it is mostly the interactions acting inside or between the bands $\alpha = 0$ and $\alpha = 2$ that correlate with $T_c$.

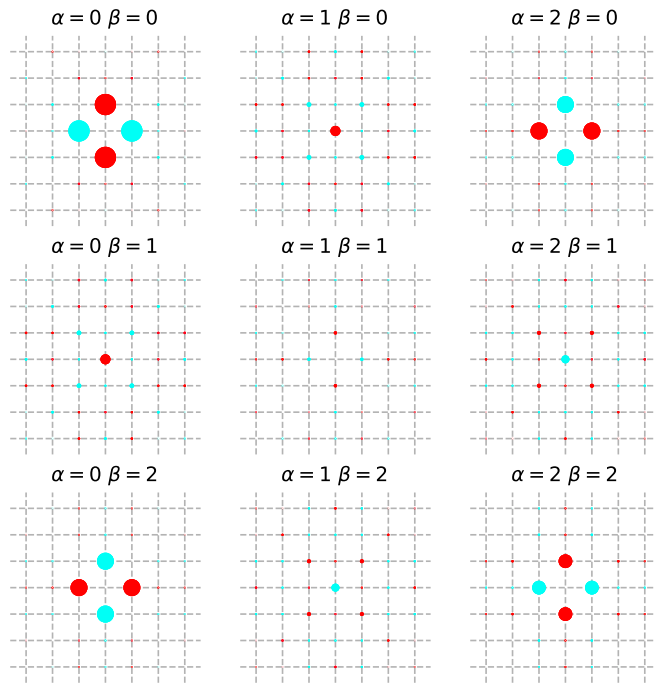

Figure 10. The correlator $\langle d^\dagger_{\alpha,\uparrow,\mathbf{r}} d^\dagger_{\beta,\downarrow,\mathbf{0}} \rangle$ under the assumption of the standard, nearest-neighbor $d$-wave pairing on the $d$-orbitals, given by Eq. 44, with $\Delta = 1$. The example is data point (4). Area of points is the amplitude, color is the phase.


\end{widetext}
\end{document}